\documentclass[aps, 10pt, prd, onecolumn, notitlepage, preprintnumbers, amsmath, amssymb, nofootinbib, superscriptaddress, a4paper]{revtex4-1}
\pdfoutput=1
\usepackage[T1]{fontenc}

\usepackage{amsthm}
\usepackage{upgreek}
\usepackage{slashed}

\usepackage[english]{babel}
\usepackage[utf8]{inputenc}
\usepackage{graphicx}
\usepackage{color}
\usepackage{amsfonts,amsthm}
\usepackage{bm,bbm}
\usepackage{xcolor}
\usepackage{lipsum}

\newcommand{\cA}{\ensuremath{\mathcal A} }
\newcommand{\cAb}{\ensuremath{\overline{\mathcal A}} }

\newcommand{\Cbb}{\ensuremath{\mathbb C} }
\newcommand{\cgrav}{\ensuremath{c_{\text{grav}}} }
\newcommand{\cD}{\ensuremath{\mathcal D} }
\newcommand{\cDb}{\ensuremath{\overline{\mathcal D}} }
\newcommand{\cF}{\ensuremath{\mathcal F} }
\newcommand{\cFb}{\ensuremath{\overline{\mathcal F}} }

\newcommand{\Ibb}{\ensuremath{\mathbb I} }
\newcommand{\cN}{\ensuremath{\mathcal N} }
\newcommand{\ve}{\mathbf{e}}
\newcommand{\vn}{\mathbf{n}}
\newcommand{\vy}{\mathbf{y}}

\newcommand{\cO}{\ensuremath{\mathcal O} }
\newcommand{\cP}{\ensuremath{\mathcal P} }
\newcommand{\cQ}{\ensuremath{\mathcal Q} }
\newcommand{\Rbb}{\ensuremath{\mathbb R} }
\newcommand{\cU}{\ensuremath{\mathcal U} }
\newcommand{\cUb}{\ensuremath{\overline{\mathcal U}} }
\newcommand{\Zbb}{\ensuremath{\mathbb Z} }
\newcommand{\xbar}{\ensuremath{\overline x} }
\newcommand{\al}{\ensuremath{\alpha} }
\newcommand{\ga}{\ensuremath{\gamma} }

\newcommand{\lam}{\ensuremath{\lambda} }

\newcommand{\lalat}{\ensuremath{\lam_{\text{lat}}} }
\newcommand{\hatbmu}{\widehat{\boldsymbol{\mu}}}

\newcommand{\glN}{\ensuremath{\mathfrak{gl}(N, \Cbb)} }
\newcommand{\nn}{\nonumber }
\newcommand{\chidof}{\ensuremath{\mbox{$\chi^2/\text{d.o.f.}$}}}
\newcommand{\pf}{\ensuremath{\text{pf}\,} }
\newcommand{\Tr}[1]{\ensuremath{\text{Tr}\left[ #1 \right]} }
\newcommand{\vev}[1]{\ensuremath{\left\langle #1 \right\rangle} }
\newcommand{\pderiv}[2]{\ensuremath{\frac{\partial #1}{\partial #2}} }
\newcommand{\fig}[1]{Fig.~\ref{#1}}
\newcommand{\tab}[1]{Table~\ref{#1}}
\newcommand{\secref}[1]{Sec.~\ref{#1}}
\newcommand{\refcite}[1]{Ref.~\cite{#1}}
\def\figheight{5.2 cm}

\usepackage[colorlinks=true, linkcolor=blue, urlcolor=blue, citecolor=blue, anchorcolor=blue]{hyperref}

\begin{document}
\title{Testing holography using lattice super-Yang--Mills on a 2-torus}

\author{Simon Catterall}
\email{smcatter@syr.edu}
\affiliation{Department of Physics, Syracuse University, Syracuse, New York 13244, United States}

\author{Raghav G.~Jha}
\email{rgjha@syr.edu}
\affiliation{Department of Physics, Syracuse University, Syracuse, New York 13244, United States}

\author{David Schaich}
\email{schaich@itp.unibe.ch}
\affiliation{Department of Physics, Syracuse University, Syracuse, New York 13244, United States}
\affiliation{AEC Institute for Theoretical Physics, University of Bern, 3012 Bern, Switzerland}

\author{Toby Wiseman}
\email{t.wiseman@imperial.ac.uk}
\affiliation{Theoretical Physics Group, Blackett Laboratory, Imperial College, London SW7 2AZ, United Kingdom}

\begin{abstract}
  We consider maximally supersymmetric SU($N$) Yang--Mills theory in Euclidean signature compactified on a flat two-dimensional torus with anti-periodic (`thermal') fermion boundary conditions imposed on one cycle.
  At large $N$, holography predicts that this theory describes certain black hole solutions in Type~IIA and IIB supergravity, and we use lattice gauge theory to test this.
  Unlike the one-dimensional quantum mechanics case where there is only the dimensionless temperature to vary, here we emphasize there are two more parameters which determine the shape of the flat torus.
  While a rectangular Euclidean torus yields a thermal interpretation, allowing for skewed tori modifies the holographic dual black hole predictions and results in another direction to test holography.
  Our lattice calculations are based on a supersymmetric formulation naturally adapted to a particular skewing.
  Using this we perform simulations up to $N = 16$ with several lattice spacings for both skewed and rectangular tori.
  We observe the two expected black hole phases with their predicted behavior, with a transition between them that is consistent with the gravity prediction based on the Gregory--Laflamme transition.
\end{abstract}

\maketitle

\section{Introduction}
Maximally supersymmetric Yang--Mills (SYM) theory in $p + 1$ dimensions has been conjectured to provide a holographic description of string theories containing D$p$-branes.
Specifically, this gauge/gravity duality states that ($p + 1$)-dimensional SYM with gauge group SU($N$) is dual to a Type~IIA (even $p$) or Type~IIB (odd $p$) superstring containing $N$ coincident D$p$-branes in the `decoupling' limit~\cite{Itzhaki:1998dd, Aharony:1999ti}.
The $p = 3$ case corresponds to superconformal $\cN = 4$ SYM in four dimensions and yields the original AdS/CFT correspondence~\cite{Maldacena:1997re}.
In this paper we focus on the maximally supersymmetric Yang--Mills in two dimensions at finite temperature, with the spatial circle direction compactified with periodic fermion boundary conditions (BCs) about it.

In this context, at large~$N$ and low temperatures, the dual string theory is well described by supergravities whose dynamics are given by certain charged black holes.
Two classes of black hole are required to describe these dynamics---those that wrap the spatial circle (so-called `homogeneous black strings') and those that are localized on it (`localized black holes')~\cite{Susskind:1997dr, Barbon:1998cr, Li:1998jy, Martinec:1998ja, Aharony:2004ig, Aharony:2005ew}.
Indeed this system of black hole solutions is related by a simple transform to the static uncharged black holes arising in pure gravity in ten dimensions with one spatial dimension wrapped into a circle, i.e., ten-dimensional Kaluza--Klein theory~\cite{Aharony:2004ig, Harmark:2004ws} (for a review of black holes in Kaluza--Klein theory see~\cite{Horowitz:2011cq}).
The two classes have different thermodynamic behaviors, and there is a first-order Gregory--Laflamme~\cite{Gregory:1993vy} phase transition between them in the gravity dual.
According to holography, all this should be reproduced by the thermal physics of the SYM.
In particular, the phase transition is a deconfinement transition associated to the spatial circle, the magnitude of the spatial Wilson line giving an order parameter.
It is thought that this transition extends to high temperatures where an intricate phase structure has been revealed from numerical and analytic treatments~\cite{Aharony:2004ig, Kawahara:2007fn, Mandal:2009vz}.

The remarkably subtle nature of gauge/gravity duality has meant that whilst SYM thus provides a fundamental and microscopic quantum description of certain gravity systems, there still is no `proof' or derivation of this black hole thermodynamics from ($p + 1$)-dimensional SYM directly.
Indeed even understanding the local structure of the dual ten-dimensional spacetime which emerges from the strongly coupled SYM theory remains a mystery.
While there has been some heuristic analytic treatment for general $p$ that hints how certain aspects of black hole thermodynamics can be seen within the SYM theory~\cite{Smilga:2008bt, Wiseman:2013cda, Morita:2014ypa}, and an approximation scheme developed for $p = 0$~\cite{Kabat:1999hp, Kabat:2001ve, Kabat:2000zv, Lin:2013jra}, a full derivation showing the SYM reproduces dual black hole behavior remains an important challenge in quantum gravity.

With only limited success from analytic treatment, it is natural to apply lattice field theory, which is well suited to study the thermodynamics of strongly coupled systems (see for example the recent review~\cite{Hanada:2016jok}).
Starting with~\cite{Hanada:2007ti, Catterall:2007fp}, several works over the past decade have studied the thermal behavior of the $p = 0$ SYM quantum mechanics, where again gravity provides a black hole prediction to be tested, and striking agreement has been seen~\cite{Anagnostopoulos:2007fw, Catterall:2008yz, Hanada:2008gy, Hanada:2008ez, Kadoh:2015mka, Filev:2015hia, Filev:2015cmz, Hanada:2016zxj, Berkowitz:2016tyy, Berkowitz:2016jlq, Asano:2016xsf, Rinaldi:2017mjl}.
However the dual gravity in that setting is simpler than in the $p = 1$ case we focus on here, where there are different black holes to probe and a gravity phase transition to observe.
Less effort has been directed at this two-dimensional case, where the state of the art until recently was simply to provide evidence for the transition at small $N \le 4$~\cite{Catterall:2010fx}.\footnote{There have also been some noteworthy numerical studies of non-maximal $\cN = (2, 2)$ SYM~\cite{Catterall:2006jw, Suzuki:2007jt, Kanamori:2007ye, Kanamori:2007yx, Kanamori:2008bk, Catterall:2008dv, Kanamori:2008yy, Kanamori:2009dk, Hanada:2009hq, Hanada:2010qg, Catterall:2011aa, Kamata:2016xmu, Catterall:2017xox, August:2018esp} and super-QCD~\cite{Catterall:2015tta} in two dimensions.}
One of the main goals of this work is to improve the lattice study of this phase transition, working at larger~$N$ and smaller lattice spacings.
We will also provide large-$N$ tests of the detailed thermal behavior of the two different classes of black holes.
(See also the recent conference proceedings~\cite{Kadoh:2016eju, Kadoh:2017mcj} for other lattice work in this direction.)

Conventionally one studies thermal physics in the canonical ensemble by considering the Euclidean theory.
This lives on a flat rectangular 2-torus, with the spatial cycle being the circle of the original theory, and the Euclidean time cycle having anti-periodic BCs for fermions and period equal to the inverse temperature.
The path integral then plays the role of a thermal partition function.
An important point we emphasize in this work is that one may also consider the Euclidean theory on a flat but skewed torus as discussed in~\cite{Aharony:2005ew}.
This no longer corresponds to the Lorentzian theory at finite temperature, but taking anti-periodic fermion BCs about Euclidean time, it may be regarded as a generalized thermal ensemble.
The key point is that this skewing is easily accommodated in the dual gravity theory, which can be treated in the Euclidean signature, and its behavior is again given in terms of solutions that may be interpreted as generalized black holes.

Studying such skewed flat tori is natural due to the lattice SYM formulation that we employ.
Recently, much progress has been made in lattice studies of the $p = 3$ theory, $\cN = 4$ SYM, using a novel construction based on discretization of a topologically twisted form of the continuum $\cN = 4$ action.
See \refcite{Catterall:2009it} for a review of this approach.
The chief merit of this new lattice construction is that it preserves a closed subalgebra of the supersymmetries at non-zero lattice spacing.
Numerical studies of the four-dimensional theory are in progress~\cite{Catterall:2014vka, Schaich:2014pda, Catterall:2014vga, Catterall:2015ira, Schaich:2015daa, Schaich:2015ppr, Schaich:2016jus}, but are quite expensive because of the large number of degrees of freedom.
In this regard, lower-dimensional theories are more tractable and can be studied extensively at large $N$ with better control over continuum extrapolations.
These lattice constructions are based on non-hypercubic Euclidean lattices, which when made periodic are naturally adapted to skewed tori.
We dimensionally reduce an $\cN = 4$ lattice system to give a discretization of two-dimensional SU($N$) SYM on an $A_2^*$ lattice, preserving four exact supercharges at non-zero lattice spacing.
Applying appropriate BCs we then carry out calculations for $N \le 16$, large enough to see dual gravity behavior.
Varying the temporal and spatial lattice extent gives the continuum SYM on tori that may be both skewed and rectangular.
We confirm that both phases of dual black hole behavior are seen in the appropriate low-temperature regime, and we see consistency between the generalized SYM thermodynamics and that predicted by gravity.
We also see a transition between these phases, again compatible with the expectation from gravity, which extends to high temperature as expected.

The plan of the paper is as follows.
In \secref{sec:gravity} we review the known predictions for large-$N$ thermal two-dimensional SYM on a spatial circle---i.e., SYM on a flat rectangular Euclidean 2-torus.
Then in \secref{sec:skewed} we discuss how this picture generalizes for a flat skewed Euclidean 2-torus.
In \secref{sec:lattice} we present our lattice construction for this skewed continuum theory.
Then in \secref{sec:results} we discuss our results, focusing on how the various gravity predictions are confirmed.
We end the paper with a brief discussion.

\section{\label{sec:gravity}Review of thermal large-$N$ $(1 + 1)$-dimensional SYM on a circle}
We now review the predictions for large-$N$ $p = 1$ SYM, compactified on a circle of size $L$ at temperature $T = 1 / \beta$, derived in various limits and using input from the dual gravity theory~\cite{Itzhaki:1998dd, Li:1998jy, Martinec:1998ja, Aharony:2004ig, Aharony:2005ew, Mandal:2011hb}.
We treat the thermal theory in Euclidean signature, with Euclidean time $\tau \sim \tau + \beta$, so that it lives on a flat rectangular 2-torus, with side lengths $\beta$ and $L$.
Fermions have thermal (anti-periodic) BCs on the Euclidean time circle, and are taken periodic on the spatial circle.
Starting in \secref{sec:skewed} we will consider the theory on a \emph{skewed} torus.
However, it will be useful to review the rectangular torus case first, as the skewed case will be similar.

The Euclidean action of the theory is
\begin{align}
  S & = S_{\text{Bos}} + S_{\text{Ferm}} \cr
  S_{\text{Bos}}  & = \frac{N}{\lam} \int d\tau\, dx\ \Tr{\frac{1}{4}F_{\mu\nu}F^{\mu\nu} + \frac{1}{2}\left(D_{\mu} X^I\right)^2 - \frac{1}{4}\left[X^I, X^J\right]^2} \cr
  S_{\text{Ferm}} & = \frac{N}{4\lam} \int d\tau\, dx\ \text{Tr}\bigg[\Psi \left(\slashed{D} - \left[\Gamma^I X^I, \, \cdot \,\right]\right) \Psi\bigg].       \label{eq:SYMaction}
\end{align}
Here $X^I$ with $I = 2, \ldots, 9$ are the eight spacetime scalars representing the transverse degrees of freedom of the branes.
They are $N\times N$ hermitian matrices in the adjoint representation of the gauge group.
The fermion $\Psi$ and matrices $\Gamma^I$ descend from a dimensional reduction of a ten-dimensional Euclidean Majorana--Weyl spinor, with $\Psi$ also transforming in the adjoint.
The dimensionful 't~Hooft coupling $\lam = N g_{YM}^2$ may be used to construct two dimensionless quantities that control the dynamics: $r_{\beta} = \beta \sqrt{\lam}$ and $r_L = L \sqrt{\lam}$.
We define the dimensionless temperature $t = 1 / r_{\beta}$.
Since we are interested in the large-$N$ 't~Hooft limit we wish to consider $N \to \infty$ with $r_{\beta}$ and $r_L$ fixed.
The main observables we consider are thermodynamic quantities related to the expectation value of the bosonic action, and also the Wilson loop magnitudes $P_{\beta}$ and $P_L$ (normalized to 1),\footnote{We define the Wilson loop to be the trace of the Wilson line $\cP e^{i\int_{\beta, L} A}$ around a closed path.} where
\begin{equation}
  \label{eq:Pdefn}
  P_{\beta, L} = \frac{1}{N} \vev{\left| \Tr{\cP e^{i\oint_{\beta, L} A}} \right|},
\end{equation}
which wrap about the Euclidean thermal circle and spatial circle of the two-dimensional Euclidean torus, respectively.
For the large-$N$ theory these act as order parameters for phase transitions associated to breaking of the $Z_N$ center symmetry of the gauge group.\footnote{Since we are at finite volume we can only have a phase transition at large $N$.}
For the thermal circle this is the usual thermal deconfinement transition, with vanishing Polyakov loop $P_{\beta} = 0$ at large~$N$ indicating the (unbroken) confined phase, and $P_{\beta} \ne 0$ being the (broken) deconfined phase.
We will use similar terminology for $P_L$, namely that $P_L \ne 0$ indicates `deconfined' spatial behavior while $P_L = 0$ corresponds to `confined' spatial behavior.

\subsection{\label{sec:rectHighTemp}High-temperature limit}
Consider the high-temperature limit of the SYM~\cite{Aharony:2004ig, Aharony:2005ew}.
Then when $r_{\beta}^3 \ll r_L$ we may integrate out Kaluza--Klein modes on the thermal circle and reduce to a bosonic quantum mechanics (BQM) consisting of the zero modes on the thermal circle.
Due to the thermal fermion BCs, this is now the bosonic truncation of the $p = 0$ SYM, as the fermions are projected out in the reduction.
Now the 't~Hooft coupling $\lam_{\text{BQM}}$ is related to the original two-dimensional coupling as $\lam_{\text{BQM}} = \frac{\lam}{\beta}$ and the dynamics implies $\oint_{\beta} A \sim 0$ so that $P_{\beta} \ne 0$ indicating thermal deconfinement.

Taking the small-volume limit, $L^3 \lam_{\text{BQM}} \ll 1$, the dynamics of this BQM (and hence the full SYM) is governed by a bosonic matrix integral of scalar and gauge field zero modes.
These dynamics imply that $\oint_L A \sim 0$, so that the SYM theory in this regime is spatially deconfined with $P_L \ne 0$.
Following Refs.~\cite{Hotta:1998en, Kawahara:2007ib} the leading behavior of the BQM energy in this regime goes as $E_{\text{BQM}} \simeq 6N^2 / L$.
The action behaves as $\vev{S_{\text{BQM}}} = -E_{\text{BQM}} L / 3$ (see, e.g.,~\cite{Catterall:2007fp}).
We expect the BQM action to give the SYM bosonic action, $S_{\text{Bos}}$, when the BQM describes it, since the fermions are decoupled in this limit.
Hence we expect the SYM bosonic action to go as $\vev{S_{\text{Bos}}} \simeq -2N^2$.
Since this limit applies when we have integrated out both the temporal and spatial Kaluza--Klein modes, reducing to only a bosonic matrix integral, its behavior is common to any high-temperature, small-volume limit, $r_L, r_{\beta} \ll 1$.

For finite volume, $L^3 \lam_{\text{BQM}} \sim 1$, this BQM has an interesting dynamics at large $N$.
This has been studied numerically and analytically in~\cite{Aharony:2004ig, Kawahara:2007fn, Mandal:2009vz, Azuma:2014cfa} with the conclusion that there is a deconfinement transition around $L^3 \lam_{\text{BQM}} \simeq 1.4$.
However the order of the transition is difficult to determine~\cite{Aharony:2004ig}.
Either it is a first-order transition (as most recently found in~\cite{Azuma:2014cfa}) or it is a strong second-order transition (as discussed in the earlier~\cite{Kawahara:2007fn, Mandal:2009vz}), in which case there is another very close-by third-order Gross--Witten--Wadia (GWW)~\cite{Gross:1980he, Wadia:1980cp} transition as well.

\subsection{\label{sec:dualGrav}Dual gravity for low temperatures}
At large~$N$ and low temperatures $r_{\beta} \gg 1$, holography predicts a gravity dual given by D1-charged black holes in IIB supergravity~\cite{Itzhaki:1998dd}.
These have a simple solution, with Euclidean string frame metric and dilaton given as
\begin{align}
  \label{eq:IIBmetric}
  ds^2_{\text{IIB, string}} & = \al' \left(\frac{U^3}{\sqrt{d_1 \lam}}\left[\left(1 - \frac{U_0^6}{U^6} \right) d\tau^2 + dx^2\right] + \frac{\sqrt{d_1 \lam}}{U^3} \left[\frac{dU^2}{1 - \frac{U_0^6}{U^6}} + U^2 d\Omega^2_{(7)}\right]\right) \cr
  e^\phi & = 2\pi \frac{\lam}{N} \frac{\sqrt{d_1 \lam}}{U^3},
\end{align}
where $d_1 = 2^6 \pi^3$ and $U_0^2 = \frac{2 \pi \sqrt{d_1 \lam}}{3 \beta}$.
There is a 2-form potential yielding $N$ units of D1~charge, and the spatial circle $x$ corresponds to that in the SYM, with $x \sim x + L$.
Large~$N$ is required to suppress string quantum corrections to the supergravity.
In order to suppress $\al'$ corrections we require $1 \ll r_{\beta}$, and to avoid winding mode corrections about the circle we need $r_{\beta} \ll r_L^2$.

When $r_{\beta} \sim r_L^2$ one indeed finds that this solution is unstable to stringy winding modes on the spatial circle $x$~\cite{Susskind:1997dr, Barbon:1998cr, Li:1998jy, Martinec:1998ja, Aharony:2004ig, Aharony:2005ew}.
This is seen by passing to a second gravity dual by T-dualizing on this circle direction to obtain a solution in IIA supergravity, which in string frame is
\begin{align}
  \label{eq:IIAmetric}
  ds^2_{\text{IIA, string}} & = \al' \left(\frac{U^3}{\sqrt{d_1 \lam}}\left[\left(1 - \frac{U_0^6}{U^6}\right) d\tau^2\right] + \frac{\sqrt{d_1 \lam}}{U^3} \left[\frac{dU^2}{1 - \frac{U_0^6}{U^6}} + U^2 d\Omega^2_{(7)} + d\bar{x}^2\right]\right) \cr
  e^\phi & = (2\pi)^2 \frac{\lam}{N} \left(\frac{\sqrt{d_1 \lam}}{U^3}\right)^{\frac{3}{2}}.
\end{align}
Now the spatial coordinate $\bar{x} \sim \bar{x} + L_{\text{IIA}}$ is compact, but due to the T-duality, has period $L_{\text{IIA}} = (2\pi)^2 \al' / L$, and there is a 1-form potential supporting D0~charge.
The D1~charge of Eq.~\eqref{eq:IIBmetric} is given as a distribution of D0~charge smeared homogeneously over the circle.
This gravity solution is a good dual for the SYM at large~$N$ and $1 \ll r_{\beta}$ (to suppress string quantum and $\al'$ corrections, respectively).
To avoid winding mode corrections about the circle we also require $r_L \ll r_{\beta}$.
In particular, for $1 \ll r_{\beta}$ this T-dual frame overlaps the regime $r_L \ll r_{\beta} \ll r_L^2$ where the IIB dual exists and describes the physics.
It describes the regime where the IIB solution fails and becomes unstable to winding modes, $r_{\beta} \sim r_L^2$, and also covers smaller circle sizes all the way down to the limit $r_L \to 0$ where the physics is that of the dimensionally reduced SYM quantum mechanics.

The above solution is homogeneous on the circle---a `homogeneous black string'.
The black hole horizon wraps over the circle direction and has a cylindrical topology $\Rbb \times S^7$.
Being related by T-duality it has precisely the same thermodynamics as the IIB solution above.
Namely it predicts the thermodynamic behavior
\begin{equation}
  \label{eq:D1phase}
  \frac{f_{\text{homog}}}{N^2 \lam} = -\frac{2^4 \pi^{\frac{5}{2}}}{3^4} t^3 \simeq -3.455 t^3
\end{equation}
for the SYM free energy density $f$, with $t = 1 / r_{\beta}$ the dimensionless temperature.
However what was a winding mode in the original frame is now a classical Gregory--Laflamme (GL) instability in this IIA frame.
One finds the above solution is dynamically unstable when
\begin{align}
  r_L^2 & \le c_{\text{GL}} r_{\beta} &
  c_{\text{GL}} & \simeq 2.24,
\end{align}
where the constant is determined by numerically solving the differential equation that governs the marginal instability mode~\cite{Aharony:2004ig, Harmark:2004ws}.

Thus at smaller circle sizes the above solution remains, but it is not the relevant one for the dynamics, which instead is given in terms of a `localized black hole' solution.
This is inhomogeneous over the circle direction, with the black hole horizon being localized on the circle and having a spherical topology $S^8$.
From a gravitational perspective the parameter that is varied to move between different solutions is the dimensionless ratio of the size of the horizon as compared to the size of the spatial IIA circle, $L_{\text{IIA}}$.
Translating to our SYM variables, this is proportional to $r_L^2 / r_{\beta}$.

These localized black hole solutions are not known analytically.\footnote{These localized solutions are considerably more complicated than the homogeneous ones as the metric and matter fields have explicit dependence on the circle direction \xbar as well as on the radial direction $U$.  Hence to find the solutions one must solve partial differential equations rather than the ordinary differential equations of the homogeneous case that depends only on $U$.}
However, recently the challenging numerical construction of these solutions has been performed~\cite{Dias:2017uyv}.
Following expectations, \refcite{Dias:2017uyv} found that for large $r_L^2 / r_{\beta}$ the thermodynamic behavior is dominated by the homogeneous phase.
At
\begin{align}
  \label{eq:cgrav}
  r_L^2 & = \cgrav r_{\beta} &
  \cgrav & \simeq 2.45
\end{align}
there is a first-order phase transition to the localized phase, which then dominates the homogeneous one for smaller $r_L^2 / r_{\beta}$, having lower free energy density.
The value of \cgrav is determined numerically, and we see it is rather close to $c_{\text{GL}}$.

While the analytic form of these localized solutions is not known generally, they do simplify in the limit that the horizon is small compared to the circle size.
In SYM variables this is the case for $y = r_L^2 / r_{\beta} \ll 1$, where the solutions have an approximate behavior~\cite{Harmark:2004ws}
\begin{equation}
  \label{eq:D0phase}
    \frac{1}{N^2 \lam} f^{\text{loc}} = -\left(\frac{2^{21}\cdot 3^2 \cdot 5^7 \pi^{14}}{7^{19}}\right)^{\frac{1}{5}} \frac{t^{\frac{14}{5}}}{r_L^{\frac{2}{5}}} \left(1 + \left(\frac{2 \cdot 3^7 \cdot 5^2}{7^{14} \pi^{21}}\right)^{\frac{1}{5}} \zeta(7) y^{\frac{14}{5}} + \cO\left(y^{\frac{28}{5}}\right)\right).
\end{equation}
Of relevance for us is that the position of the phase transition found in \refcite{Dias:2017uyv} is such that the leading term in the above approximation~\eqref{eq:D0phase} agrees very well (to the percent level) with the full numerical solutions over the full range where this phase dominates the thermodynamics.
This means that while one generally requires the numerical solutions of~\cite{Dias:2017uyv} to deduce the thermodynamics of a given localized solution, since we are only concerned with this localized branch for $r_L^2 / r_{\beta}$ where it dominates the thermodynamics we may very accurately approximate the thermal behavior for such solutions using~\eqref{eq:D0phase}.\footnote{Using only the leading term in~\eqref{eq:D0phase} and comparing with~\eqref{eq:D1phase} gives an approximation for the phase transition with \cgrav within 2\% of the numerically computed value in~\eqref{eq:cgrav}.  Including the subleading term improves this to be consistent with the value in~\eqref{eq:cgrav} within its numerical uncertainty.}
In previous lattice investigations~\cite{Catterall:2010fx} this transition was probed using small $N = 3$ and 4, finding evidence for consistency with the value for \cgrav in~\eqref{eq:cgrav}.
In this work we will improve the lattice study of the phase diagram, employing larger $N$ and smaller lattice spacings.

We will refer to the homogeneous phase as the \emph{D1~phase}, since in the IIB duality frame it is the D1-brane solution, although we note that it may also be seen as a homogeneous D0-brane solution in the IIA frame.
We will refer to the localized phase as the \emph{D0~phase}, since it may only be seen in gravity in the IIA frame where it is a localized D0-brane black hole.

Since all these gravity solutions are static black holes, their Euclidean time circle is contractible so we expect a deconfined Polyakov loop, $P_{\beta} \ne 0$.\footnote{Recall that when IIB gravity provides a good dual description of the SYM we expect the Wilson loop (normalized as in~\eqref{eq:Pdefn}) about a cycle in this boundary theory to be non-vanishing if that cycle is contractible when extended into the dual bulk (such as for a cycle about Euclidean time when a horizon exists in the bulk)~\cite{Maldacena:1998im, Witten:1998zw, Aharony:2003sx}.  Conversely if a cycle is non-contractible in the bulk, we expect the corresponding Wilson loop to vanish.  Note this picture does not hold for the spatial cycle after T-dualizing to the IIA frame.  Then, instead, the distribution of D0~charge on the spatial circle is thought to determine the eigenvalue distribution of $\cP e^{i\oint_L A}$.}
In the IIB frame, as the horizon wraps over the spatial circle for the homogeneous black string, this spatial cycle is not contractible in the bulk solution.
Hence at large $N$ we expect spatial confinement, $P_L = 0$, when this homogeneous phase describes the thermodynamics (for $1 \ll \cgrav r_{\beta} < r_L^2$), with the thermal behavior given by Eq.~\eqref{eq:D1phase}.
The homogeneity of the horizon is taken to indicate that the eigenvalues of $\cP e^{i\oint_L A}$ are uniformly distributed at large $N$.
On the other hand, upon decreasing the circle size $r_L$ at fixed $r_{\beta}$ we have a first-order transition to the localized phase with thermodynamics given by Eq.~\eqref{eq:D0phase}.
Due to the localized horizon, the D0-brane charge is compactly supported on the spatial circle, so we expect the eigenvalue distribution for $\cP e^{i\oint_L A}$ is likewise compactly supported~\cite{Kol:2002xz, Aharony:2004ig}.
This implies spatial deconfinement, $P_L \ne 0$.
The phase transition curve $r_L^2 = \cgrav r_{\beta}$ in the gravity regime, $r_{\beta} \gg 1$, therefore corresponds to a first-order spatial deconfinement transition associated to $P_L$.

We emphasize that we are interested in temperatures and circle sizes where $r_{\beta}$ and $r_L \sim \cO(N^0)$ in the large-$N$ limit.
If we were to take ultralow temperatures $r_{\beta} \to \infty$ as some sufficiently large positive power of $N$, then the gravity predictions above would cease to be valid because the gravity would become strongly coupled near the black hole horizons.
In particular, for $r_{\beta} \sim N$ the theory is thought to enter a conformal phase described by a free orbifold CFT~\cite{Itzhaki:1998dd, Aharony:1999ti}, which we will not explore in this work.

\subsection{Summary for SYM on a rectangular torus}
\begin{figure}[tbp]
  \centering
  \includegraphics[width=0.5\linewidth]{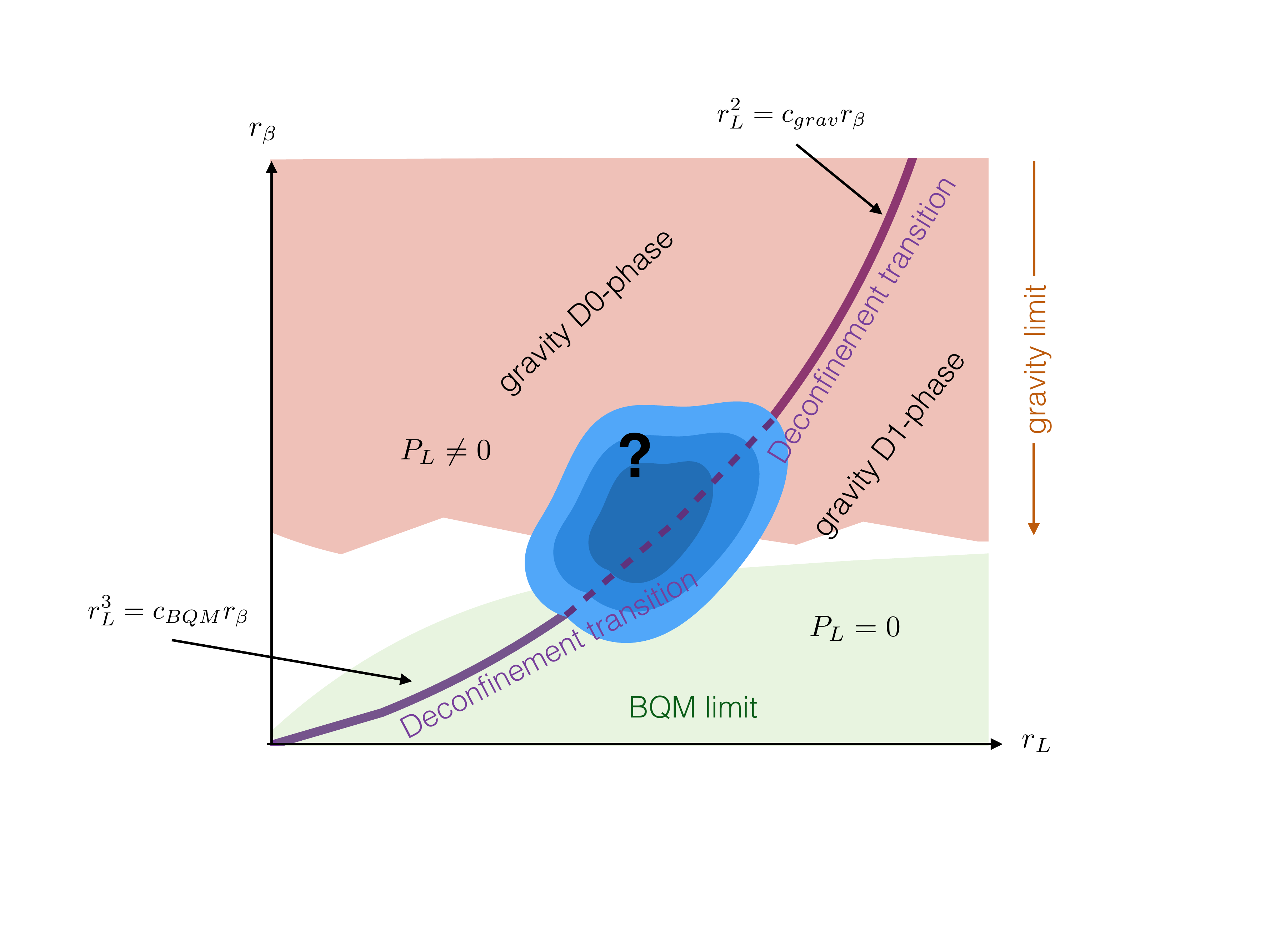} 
  \caption{\label{fig:summaryrect}Summary of the expected phase structure for the SYM theory on a rectangular 2-torus.}
\end{figure}

For large-$N$ two-dimensional SYM on a \emph{rectangular} Euclidean 2-torus we have two dimensionless parameters $r_{\beta}$ and $r_L$.
Assuming that $r_{\beta}, r_L \sim \cO(N^0)$ in the large-$N$ limit we have the following expectations:
\begin{itemize}
  \item The high-temperature, small-volume limit $r_{\beta}, r_L \ll 1$ is described by the dynamics of scalar and gauge zero modes.
        We expect $P_{\beta}, P_L \ne 0$ and $\vev{S_{\text{Bos}}} \simeq -2N^2$.
  \item The high-temperature limit $r_{\beta}^3 \ll r_L$ reduces to BQM.
        Here we expect $P_{\beta} \ne 0$.
        For $r_L^3 < c_{\text{BQM}} r_{\beta}$ with $c_{\text{BQM}} \simeq 1.4$ we expect $P_L \ne 0$, with a deconfinement transition to $P_L = 0$ for $r_L^3 > c_{\text{BQM}} r_{\beta}$.
  \item The low-temperature limit, $r_{\beta} \gg 1$, admits a gravity-dual black hole description, so $P_{\beta} \ne 0$, with the free energy density depending on the ratio $r_L^2 / r_{\beta}$.
        For $r_L^2 = \cgrav r_{\beta}$, with $\cgrav = 2.45$ there is a first-order deconfinement phase transition with respect to $P_L$.
        The D1~phase, for $r_L^2 > \cgrav r_{\beta}$, has free energy density given by Eq.~\eqref{eq:D1phase} and $P_L = 0$.
        The D0~phase, for $r_L^2 < \cgrav r_{\beta}$, has $P_L \ne 0$ with the free energy density well approximated by Eq.~\eqref{eq:D0phase}.
\end{itemize}
This is illustrated in \fig{fig:summaryrect}.

\section{\label{sec:skewed}Behaviour on a skewed torus}
We now discuss what happens to the Euclidean theory when it is placed on a skewed torus.
The motivation is twofold.
First we emphasize that skewing the torus provides a new direction to deform the theory and hence a new independent test of holography, given again that gravity dual predictions exist.
Second, modern supersymmetric lattice constructions often naturally live on non-hypercubic lattices, which in turn are naturally adapted to giving a continuum Euclidean theory on a skewed torus.

We take the SYM to live on the flat 2-torus generated as a quotient of the two-dimensional plane.
Writing the metric as
\begin{equation}
  ds^2_{T^2} = d\tau^2 + dx^2,
\end{equation}
this is generated by the identifications
\begin{align}
  (\tau, x) & \sim (\tau, x) + \vec{\beta} & & \text{fermions anti-periodic} \\
  (\tau, x) & \sim (\tau, x) + \vec L & & \text{fermions periodic}
\end{align}
for vectors $\vec{\beta}$ ($\vec L$) about the thermal (spatial) circles with anti-periodic (periodic) fermion BCs.
These vectors defining the cycles have lengths and dot product
\begin{align}
  \beta & = |\vec{\beta}| &
  L & = |\vec L| &
  \ga & = \frac{\vec{\beta} \cdot \vec L}{\beta L}
\end{align}
with $|\ga| < 1$ to ensure a non-degenerate torus.
Thus the geometry is determined by three parameters---in addition to the $\beta$ and $L$ of the rectangular case there is the dimensionless skewing parameter $\ga$.
For any value of \ga (not just the rectangular case $\ga = 0$) we will be able to compare our numerical SYM results to gravity predictions, and hence test holography.

We have constructed the torus by a quotient of the plane by the vectors $\vec{L}$ and $\vec{\beta}$.
However any SL$(2, \Zbb)$ transformation of these vectors will define the \emph{same} geometric torus.
The anti-periodic BCs about the $\beta$ circle, and periodic ones about $L$, complicate this slightly.
Consider the transformation generated by the following subgroup of SL$(2, \Zbb)$:
\begin{align}
  \label{eq:SLtransform}
  \left(\begin{array}{c}
    \vec{L}' \\
    \vec{\beta}'
  \end{array}\right) & = M \cdot \left(\begin{array}{c}
    \vec{L} \\
    \vec{\beta}
  \end{array}\right) &
  M & = \left(\begin{array}{cc}
    a & 2n \\
    c & 2m - 1
  \end{array}\right) \in \text{SL}(2, \Zbb), &
  n, m, c & \in \Zbb,
\end{align}
where we note that then $a \in 2\Zbb - 1$.
Then the 2-torus defined by $\vec{\beta}'$ (anti-periodic fermions) and $\vec{L}'$ (periodic fermions) is the \emph{same} as that defined by $\vec{\beta}$ and $\vec{L}$.
The fermion BCs restrict us to a subgroup of the full SL action, so that the new $\vec{\beta}'$ has an odd coefficient multiplying $\vec{\beta}$ to maintain anti-periodicity, and likewise $\vec{L}'$ has an even coefficient multiplying $\vec{\beta}$.

In the rectangular case, we have a Lorentzian thermal interpretation of the physics, where we identify $\beta$ as inverse temperature in a canonical ensemble.
In the skewed case this is no longer true.
Nonetheless, we may regard this case as a generalized thermal ensemble, with $1 / \beta$ playing the role of generalized temperature.
As for a rectangular torus we again work with dimensionless
\begin{align}
  r_{\beta} & = \beta \sqrt{\lam} &
  r_L & = L \sqrt{\lam},
\end{align}
now supplemented by the dimensionless parameter $\ga$.
We denote $t = 1 / r_{\beta}$ the generalized dimensionless temperature, and also define the aspect ratio
\begin{equation}
  \al = \frac{L}{\beta} = \frac{r_L}{r_{\beta}}.
\end{equation}
The 2-volume is then given as $\text{Vol}_{T^2} = \beta^2 \al \sqrt{1 - \ga^2}$.
In practice on the lattice it will be convenient to scan the parameter space by fixing the `shape' of the torus set by $(\al, \ga)$ and varying the dimensionless temperature $t$ that controls the size of the torus in units of the SYM coupling $\lam$.

The redundancy in our description of the torus using $\vec{\beta}$ and $\vec{L}$ given in~\eqref{eq:SLtransform} translates into an invariance in this parameterization: a set $\al, \ga, t$ defined from $\vec{\beta}$ and $\vec{L}$ is equivalent to other parameters $\al', \ga', t'$ similarly defined from $\vec{\beta}'$ and $\vec{L}'$.
In the usual manner we may describe this using the complex `modular parameter' $\uptau$, given by\footnote{Since `tau' is conventionally used for both the torus modular parameter and Euclidean time, we attempt to avoid potential confusion by using the symbols $\uptau$ for the modular parameter and $\tau$ for Euclidean time.}
\begin{equation}
  \uptau = \frac{L_{\tau} + iL_x}{\beta_{\tau} + i\beta_x} = \al \left(\ga + i\sqrt{1 - \ga^2}\right),
\end{equation}
which encodes the (dimensionless) shape parameters $\al, \ga$, and is independent of the torus scale.
The parameter transforms under the action~\eqref{eq:SLtransform} as
\begin{equation}
  \uptau' = \frac{a \uptau + 2n}{c \uptau + 2m - 1},
\end{equation}
so that $m, n, c \in \Zbb$ and $a(2m - 1) - 2nc = 1$.
We term this a \emph{restricted} modular transformation---it is a usual modular transformation, but restricted to preserve our fermion BCs (anti-periodic on the $\vec{\beta}$ cycle and periodic on the $\vec{L}$ cycle).
As for the usual modular invariance of the torus, we may define a fundamental domain for this $\uptau$ parameter under the action of this restricted transform~\eqref{eq:SLtransform}, which gives the set of inequivalent tori (taking into account fermion BCs).
A modular parameter outside this domain can then be mapped back into it using the appropriate~\eqref{eq:SLtransform}.
The usual modular transformations are generated by $\uptau \to \uptau + 1$ and $\uptau \to -1 / \uptau$, or equivalently $\uptau \to \uptau + 1$ and $\uptau \to \frac{\uptau}{\uptau + 1}$.
We provide some review of this in Appendix~\ref{app:modular}.
However for our restricted transform instead we may use the generators $\uptau \to \uptau + 2$ and $\uptau \to \frac{\uptau}{\uptau + 1}$.
The fundamental domain $D$ may then be taken to be
\begin{equation}
  D = \left\{\uptau\big|1 \le |\uptau \pm 1|, \; |\mathrm{Re}(\uptau)| \le 1\right\}.
\end{equation}
These assertions are proved in Appendix~\ref{app:modular}.
The lattice construction we use later will give torus geometries with a particular \ga (which we will see is $\ga = -1 / 2$), and we will vary the shape parameter \al of the torus and its size $t$.
Some of the torus shapes we study will correspond to $\uptau$ within the fundamental domain, and others will not.
We will generally present results in terms of \al and $t$ for this common value of $\ga$, but as we discuss later, for the shapes that fall outside the fundamental domain there is an alternate description with new $\al', \ga', t'$.\footnote{It is worth emphasizing that the way the modular parameter $\uptau$ arises is a little different to that in the familiar two-dimensional CFT setting.  In the context of Euclidean two-dimensional CFT \emph{any} 2-torus is Weyl equivalent to a flat torus with modular parameter $\uptau$ (i.e., one constructed as a quotient of the two-dimensional plane) and unit volume.  Hence for \emph{any} 2-torus the CFT partition function \emph{only} depends on the geometry through $\uptau$.  However in our case the SYM is not a CFT---in particular our theory explicitly depends on the size of the torus, and will also depend on the details of its real geometry.  We restrict ourselves to only consider 2-tori constructed as quotients of the plane, which are flat but with a skewing parameterized by $\uptau$.  Hence our SYM depends on $\uptau$ and the scale (through, say, $\beta$) because we have restricted to flat tori.  If we also added scalar curvature the situation would be more complicated (unlike for a CFT).  However in either case (CFT or our SYM) the dependence is on $\uptau$ up to modular invariance, so we may always choose $\uptau$ to be in the fundamental domain.  This is simply because the description of the torus has this invariance, and it is \emph{not} due to any symmetry properties of the field theory.}

An important aim of our lattice calculations is to see the detailed generalized thermodynamic behavior predicted by the dual black holes.
However in a skewed setting we are no longer in a strict thermal context and so potentials such as free energy are not strictly meaningful.
Instead the relevant quantity is the (logarithm of the) partition function
\begin{equation}
  \label{eq:partfunc}
  Z[t, \al, \ga] = \int D(\text{fields}) \, e^{-S}
\end{equation}
for the Euclidean action given by Eq.~\eqref{eq:SYMaction}, but now on the skewed 2-torus.
In a lattice calculation it is hard to determine the value of this integral directly, so instead we focus on expectation values, which are much more convenient to compute.
A natural observable is the expectation value of the Euclidean action $S$.
Since the fermionic part of the action is gaussian, its value is simply a constant.
Therefore we will focus on measuring the expectation value of the bosonic part of the action, $S_{\text{Bos}}$, given in Eq.~\eqref{eq:SYMaction}.
We will find it convenient to work with the average bosonic action \emph{density}, defined from the (renormalized) vev of $S_{\text{Bos}}$ in the obvious way,
\begin{equation}
  \vev{s_{\text{Bos}}} = \frac{1}{\text{Vol}_{T^2}} \vev{S_{\text{Bos}}}.
\end{equation}
We now show that $s_{\text{Bos}}$ is related to a derivative of the partition function $\ln Z$.
After scaling the bosonic fields and coordinates as
\begin{align}
  \tau' & = \frac{1}{\beta} \tau &
  x' & = \frac{1}{\beta} x &
  A_{\mu}' & = \beta A_{\mu} &
  X_i' & = \beta X_i,
\end{align}
the Euclidean action becomes
\begin{align}
  S_{\text{Bos}} & = \frac{N}{\beta^2 \lam} I[\al, \ga] \cr
  I[\al, \ga] & = \int_{T'^2} d\tau' \, dx' \, \Tr{\frac{1}{4} F_{\mu\nu}' F'^{\mu\nu} + \frac{1}{2} (D_{\mu}' X_i')^2 - \frac{1}{4} [X_i', X_j']^2}.
\end{align}
The 2-torus $T'^2$ is generated by the identifications
\begin{align}
  (\tau', x) & \sim (\tau' + 1, x) &
  (\tau', x) & \sim (\tau' + \ga \al, x + \al\sqrt{1 - \ga^2}),
\end{align}
the former anti-periodic and the latter periodic for the fermions.
Thus we have scaled out $\beta$.
The only explicit $\beta$ dependence is in the overall coupling, and the integral $I[\al, \ga]$ depends only on the dimensionless shape parameters \al and \ga (through the identifications above).
By suitably scaling the fermion fields the fermionic action can be chosen to have no $\beta$ dependence.
Then differentiating the partition function with respect to $\beta$, keeping \al and \ga fixed, we obtain
\begin{align}
  \left. \beta \pderiv{}{\beta} \ln Z \right|_{\al, \ga} & = \frac{1}{Z} \int D(\text{fields}) \left(\left. \beta \pderiv{}{\beta} \left(-\frac{N}{\beta^2 \lam} I[\al, \ga]\right) \right|_{\al, \ga}\right) e^{-S} = 2\vev{S_{\text{Bos}}} \\
  \vev{s_{\text{Bos}}} & = \frac{1}{2 \text{Vol}_{T^2}} \left. \beta \pderiv{}{\beta} \ln Z \right|_{\al, \ga}.                                                                                                             \label{bosonicaction}
\end{align}
Thus computing $s_{\text{Bos}}$ as a function of $t$, \al and $\ga$, which may be conveniently done on the lattice, gives the same information as that contained in the partition function.

As in the rectangular case we will be interested in the Wilson loops about the torus temporal and spatial cycles and their magnitudes $P_{\beta}$ and $P_L$, respectively.
Since there are equivalent presentations of the same torus, one can equally consider the loops $P_{\beta}'$ and $P_L'$, which correspond to cycles in the original representation that wrap multiples of the cycles generated by $\vec{\beta}$ and $\vec{L}$.\footnote{The center symmetry transformations of Wilson loops associated to the cycles $\vec{\beta}$ and $\vec{L}$ determine those of $\vec{\beta}'$ and $\vec{L}'$.  Suppose we define the holonomies $W_{\beta} = \Tr{\cP e^{i \oint_{\beta} A}}$ and $W_L = \Tr{\cP e^{i \oint_L A}}$ so that center symmetry acts on these as $W_{\beta} \to z_{\beta} W_{\beta}$ and $W_L \to z_L W_L$ for phases $z_{\beta, L}$ where $z_{\beta, L}^N = 1$.  Then consider a loop $W_{\beta}'$ associated to $\vec{\beta}'$ given by the linear combination of cycles $\vec{\beta}$ and $\vec{L}$ in Eq.~\eqref{eq:SLtransform}.  This will transform now as $W_{\beta}' \to z_L^c z_{\beta}^{2m - 1} W_{\beta}'$ and so is determined by the transformations of $W_{\beta}$ and $W_L$.  The same is true for $W_L'$.}

We now proceed to discuss the same limits as in the rectangular case in this skewed geometry.
We will find that analogous small circle reductions occur, and that generalized black holes still give gravity predictions for small generalized temperature $1 \ll r_{\beta}$.
We first discuss the dimensional reduction of the theory on a skewed torus, and then turn to a discussion of the gravity dual, giving predictions for the observable $s_{\text{Bos}}$.

\subsection{\label{sec:smallcircle}High-temperature limit}
We now take the skewing parameter \ga to be fixed, and consider the high-temperature limit, finding qualitatively similar behavior to the rectangular torus case we previously discussed.
In the small-volume, high-temperature limit $r_{\beta}, r_L \ll 1$ precisely the same reduction to the matrix integral occurs.
Thus we again expect $P_{\beta} \simeq P_L \ne 0$, with the bosonic action behaving as $\vev{S_{\text{Bos}}} / N^2 \simeq -2$ at large $N$.
Translating to the bosonic action density we then obtain
\begin{equation}
  \label{eq:HT}
  \frac{\vev{s_{\text{Bos}}}}{N^2 \lam} = -\frac{2}{\al \sqrt{1 - \ga^2}} t^2.
\end{equation}

At finite volume we may again reduce to an effective one-dimensional theory at high temperature.
The easiest way to understand this dimensional reduction is to note that skewed tori in the fixed-$\ga$, high-temperature limit become equivalent to nearly rectangular tori (again with small thermal circles) under a suitable transformation~\eqref{eq:SLtransform}.
Then we may simply dimensionally reduce this equivalent rectangular torus, and pull the result back to the original skewed torus parameterization given by $\ga$.

Thus we consider the high-temperature limit where we fix \ga and $r_L$, taking $r_{\beta} \to 0$ so that $\al \to \infty$.
We use the transform~\eqref{eq:SLtransform} with $c = 0$, $m = 0$ and $a = 1$, leaving $n \in \Zbb$, to find an equivalent torus that will be approximately rectangular.
This maps $\vec{\beta}$, $\vec{L}$ to
\begin{align}
  \vec{\beta}' & = \vec{\beta} &
  \vec{L}' & = \vec{L} + 2n \vec{\beta}.
\end{align}
This leaves the temperature unchanged, $t' = t$, and relates the shape parameters as
\begin{align}
  \al' \ga' & = \al \ga + 2n &
  \al' \sqrt{1 - \ga'^2} & = \al \sqrt{1 - \ga^2}.
\end{align}
In the $\al \to \infty$ limit we may choose $n$ appropriately (i.e., taking $-\al \ga / 2 \simeq n \in \Zbb$) to obtain an equivalent torus with $\ga' \simeq 0$ and $\al' \simeq \al \sqrt{1 - \ga^2} \to \infty$.
This equivalent torus is approximately rectangular and in the high-temperature limit, so from our previous discussion in \secref{sec:rectHighTemp} we may reduce on the thermal circle when
\begin{equation}
  \label{eq:reductime}
  \left(r_{\beta}' \right)^3 \ll r_L' \quad \implies \quad r_{\beta}^3 \ll r_L \sqrt{1 - \ga^2}.
\end{equation}
We obtain BQM on a circle size $L_{\text{BQM}} = L'$ with coupling $\lam_{\text{BQM}} = \lam / \beta'$, so that
\begin{align}
  \lam_{\text{BQM}} & = \frac{\lam}{\beta} &
  L_{\text{BQM}} & = L \sqrt{1 - \ga^2}.
\end{align}
Thus we have the same relation of the lower- and higher-dimensional coupling as in the rectangular case, but the circle size is related via a skewing-dependent factor.

Reducing on the thermal cycle $\vec{\beta}'$ implies that the Polyakov loop is $P_{\beta}' \sim 1$.
Given that $P_{\beta}' = P_{\beta}$, and this Polyakov loop is trivial, we expect thermal deconfinement, $P_{\beta} \ne 0$.
Then $P_L \simeq P_L' = P_{\text{BQM}}$, so our previous discussion of BQM implies the deconfinement transition in the limit of Eq.~\eqref{eq:reductime} is located at
\begin{equation}
  \label{eq:skewBQM}
  r_L^3 \simeq \frac{1.4}{\left(1 - \ga^2\right)^{\frac{3}{2}}} r_{\beta}.
\end{equation}
This is associated with a transition from a spatially confined phase with $P_L = 0$ to a deconfined one with $P_L \ne 0$ as $r_L^3 / r_{\beta}$ is reduced.

\subsection{The low-temperature dual gravity limit - D1 phase}
Considering the dual IIB gravity we find the same homogeneous D1-charged black hole solution as in Eq.~\eqref{eq:IIBmetric}, but now we take the two-dimensional torus in the field theory directions to be generated by the identifications
\begin{align}
  \label{eq:IIBident}
  (\tau, x) & \sim (\tau + \beta, x)                     & & \text{anti-periodic fermions} \cr
  (\tau, x) & \sim (\tau + \ga L, x + L\sqrt{1 - \ga^2}) & & \text{periodic fermions.}
\end{align}
Then asymptotically, when $U_0 \gg U$, the torus spanned by $\tau$ and $x$ has our required skewed geometry.
We also see that the relation between $U_0$ and $\beta$ is exactly the same as for Eq.~\eqref{eq:IIBmetric}, since the metric is locally the same as in the rectangular case, and the $\tau$ circle has the same period $\beta$.

We will regard this as a `generalized black hole' in the sense that for real $\beta$ and $\ga \ne 0$ its properties are not related directly to a physical Lorentzian black hole.
Nonetheless the Euclidean IIB solution exists.
The geometry of this solution differs only globally in the $x$ direction from the rectangular case, and is homogeneous in $x$.
The solution should be a good description of the IIB string theory again for large $N$ and $1 \ll r_{\beta}$.
We expect it to become unstable to a winding mode instability for $r_L^2 \sim r_{\beta}$.

Consider the expectation value of the SYM Euclidean Lagrangian density $\vev{L_E}$ predicted by this solution, which will be homogeneous in both $\tau$ and $x$.
Then $-\ln Z = \vev{S} = \text{Vol}_{T^2} \vev{L_E}$ when this black hole is the dominant saddle point of the path integral.
Due to the homogeneity in $x$ this Lagrangian density is only a function of $\beta$, and has no dependence on the other parameters \al and \ga determining the shape of the torus.
In the rectangular case it simply equals the free energy density of the solution, and thus generally is given by Eq.~\eqref{eq:D1phase}, so $\vev{L_E} = -N^2 \lam \cdot 2^4 \pi^{\frac{5}{2}} t^3 / 3^4$.
As before we refer to this homogeneous black string phase as the \emph{D1~phase}.
Hence if this gravitational solution dominates the partition function the theory is in the D1~phase and the bosonic action density is
\begin{equation}
  \label{eq:skewD1phase}
  \text{D1~phase:} \quad \frac{s_{\text{Bos, D1}}}{N^2 \lam} = -\frac{1}{2 \text{Vol}_{T^2}} \left. \beta \pderiv{}{\beta} \left(\text{Vol}_{T^2} \vev{L_E}\right) \right|_{\al, \ga} = -\frac{2^3 \pi^{\frac{5}{2}}}{3^4} t^3 \simeq -1.728t^3.
\end{equation}
We see explicitly that our observable doesn't depend on the skewing of the torus.
As in the rectangular case, $\partial / \partial \tau$ generates a contractible cycle due to the horizon.
Now the Euclidean time cycle of the torus is simply generated by $\partial / \partial \tau$.
The spatial cycle of the torus is now generated by $\sqrt{1 - \ga^2} \partial / \partial x + \ga \partial / \partial \tau$, and since $\partial / \partial x$ is not contractible, neither is this spatial torus cycle.
Thus when Eq.~\eqref{eq:IIBmetric} is the dominant bulk solution, this implies that the Polyakov loop $P_{\beta} \ne 0$, whereas $P_L = 0$.
Hence the dual SYM is thermally deconfined, but in a spatially confined phase.

There is an important subtlety in the above discussion: one must be careful whether the solution above~\eqref{eq:IIBmetric} does dominate, due to there being other related gravitational dual black holes~\cite{Aharony:2005ew}.
In Eq.~\eqref{eq:IIBident} we have identified with $\vec{\beta}$ and $\vec{L}$ to generate the 2-torus, but as discussed in \refcite{Aharony:2005ew} we may equally well use any equivalent pair under the transformation~\eqref{eq:SLtransform}, since they will give the same flat 2-torus asymptotically.
This will yield another inequivalent gravitational dual solution.
However, it is the pair of vectors whose corresponding modular parameter $\uptau$ lies in the fundamental domain that gives the dominant gravitational dual solution.
This is understood as follows.
Take a $\uptau$ in the fundamental domain, $D$, and a temperature $t$.
Then the Euclidean action density is as in Eq.~\eqref{eq:skewD1phase}.
Now transforming this to an equivalent $\uptau'$ outside the fundamental domain results in a new dimensionless temperature $t'$ given in terms of $t$, $\uptau$ and the transformation.
This $t'$ is lower than the fundamental-domain temperature $t$ since\footnote{Eq.~\eqref{eq:treln} may be derived neatly by noting the 2-torus volume $\beta^2 \mathrm{Im}(\uptau)$ is invariant under modular transformations.}
\begin{equation}
  \label{eq:treln}
  \left(\frac{t'}{t}\right)^2 = \frac{\mathrm{Im}(\uptau')}{\mathrm{Im}(\uptau)},
\end{equation}
and from the corollary in Appendix~\ref{app:modular} we have $\mathrm{Im}(\uptau') / \mathrm{Im}(\uptau) \le 1$ for $\uptau \in D$.
Thus we see $t' \le t$.
Hence the action density of these other gravitational saddle points, which is given by~\eqref{eq:skewD1phase} with $t \to t'$, is more positive.
Since the volume of the torus is preserved under the modular transformation~\eqref{eq:SLtransform}, then the Euclidean action is also more positive.
Thus these other dual solutions outside the fundamental domain are not the relevant saddle points to determine the partition function behavior.

Hence the subtlety is that the D1-phase prediction is given by Eq.~\eqref{eq:skewD1phase} for $\al, \ga, t$ corresponding to a $\uptau$ in the fundamental domain.
If one has a set of parameters outside the fundamental domain, one must first map them to new parameters $\al', \ga', t'$ in the fundamental domain, and then apply the formula~\eqref{eq:skewD1phase} with $t \to t'$.\footnote{The expression \eqref{eq:skewD1phase} could not hold for general $\al, \ga, t$ as it would not respect the modular invariance that the SYM on this flat torus must enjoy.  It is also worth emphasizing that one cannot naively compare Eqs.~\eqref{eq:skewD1phase} and \eqref{eq:skewD0phase} to deduce a critical temperature, since $\uptau \not\in D$ when the latter is valid.}
In this canonical representation (i.e., $\uptau' \in D$) we will have the prediction $P_{\beta}' \ne 0$ and $P_L' = 0$ for the Wilson loops.
A spatial cycle of the torus is always non-contractible, so we should have $P_L = 0$ in any equivalent representation.
However, while $P_{\beta}' \ne 0$ for the thermal cycle in the fundamental domain representation, if in the equivalent representation $\vec{\beta}$ is a linear combination of both $\vec{\beta}'$ and $\vec{L}'$ it may correspond to a non-contractible cycle in the gravity dual, so that also $P_{\beta} = 0$.
We emphasize that this does \emph{not} imply temporal confinement, since there is \emph{some} temporal loop (associated to the cycle $\vec{\beta}'$) where $P_{\beta}' \ne 0$.

\subsection{The low-temperature dual gravity limit - D0 phase}
Considering the D1~phase with $r_{\beta} \gg 1$ where the gravity is a valid description, fixing \ga and reducing the circle size to $r_L^2 \sim r_{\beta}$ we again expect winding modes on the spatial circle to become important near the horizon, as in \secref{sec:dualGrav}.
This limit however is straightforward to understand, as it implies $r_L \ll r_{\beta}$ and thus we may play a similar trick as for the small-thermal-circle dimensional reduction in \secref{sec:smallcircle}, again mapping to an equivalent almost-rectangular representation.
We set $a = 1$, $n = 0$ and $m = 0$ in the transform~\eqref{eq:SLtransform}, leaving $c \in \Zbb$.
Then
\begin{align}
  \vec{\beta}' & = \vec{\beta} + c \vec{L} &
  \vec{L}' & = \vec{L},
\end{align}
so $L' = L$ and the equivalence relates the shape parameters as
\begin{align}
  \label{eq:circred}
  \frac{1}{\al'} \ga' & = \frac{1}{\al} \ga + c &
  \frac{1}{\al'} \sqrt{1 - \ga'^2} & = \frac{1}{\al} \sqrt{1 - \ga^2}.
\end{align}
Then by choosing $c$ appropriately (i.e., taking $-\ga / \al \simeq c \in \Zbb$) in the $\al \to 0$ limit we obtain an equivalent torus with $\ga' \simeq 0$ and $\al' = \al / \sqrt{1 - \ga^2} \to 0$.

Thus for fixed non-zero \ga and $r_L \ll r_{\beta}$, so that the modular parameter $\uptau$ is far outside the fundamental domain, the above transform maps to an approximately rectangular torus with $r_L' = r_L$ and $r_{\beta}' = r_{\beta} \sqrt{1 - \ga^2}$.
For the rectangular torus we know that the phase transition from the D0~phase to the D1~phase occurs for $r_L'^2 = \cgrav r_{\beta}'$ with $\cgrav = 2.45$.
Mapping this back to our non-rectangular parameterization, we expect a transition at
\begin{equation}
  \label{eq:skewtransition}
  r_L^2 = \cgrav \sqrt{1 - \ga^2} r_{\beta}.
\end{equation}
Then for this rectangular torus we may use the approximation~\eqref{eq:D0phase} for the thermal behavior, replacing $t \to t'$.
From this we may compute $\ln Z$.
Translating back to our original temperature $t$ we may compute the bosonic action density using Eq.~\eqref{bosonicaction} to obtain, in our original non-rectangular parameterization,
\begin{equation}
  \label{eq:skewD0phase}
  \frac{s_{\text{Bos}}}{N^2 \lam} = -\left(\frac{2^{21} \cdot 3^7 \cdot 5^2 \pi^{14}}{7^{19}}\right)^{\frac{1}{5}} \frac{t^{\frac{16}{5}}}{\al^{\frac{2}{5}} \left(1 - \ga^2\right)^{\frac{7}{5}}} \left[1 - \left(\frac{2^{11} \cdot 3^2 \cdot 5^2}{7^{14} \pi^{21} \left(1 - \ga^2\right)^7}\right)^{\frac{1}{5}} \zeta(7) \left(\frac{\al^2}{t}\right)^{\frac{14}{5}} + \cO\left(\left(\frac{\al^2}{t}\right)^{\frac{28}{5}}\right)\right].
\end{equation}
In the rectangular representation when this D0~phase dominates we expect to have $P_{\beta}', P_L' \ne 0$, so this phase is both spatially and thermally deconfined.
Furthermore, since $\vec{L}' = \vec{L}$ we also have $P_L \ne 0$.
Thus $P_L$ remains an order parameter for the transition between the gravity D1 and D0~phases.

\subsection{\label{sec:summaryskew}Summary for SYM on a skewed torus}
For large-$N$ SYM on a skewed torus with fixed $\ga$, upon varying $r_L$ and $r_{\beta}$ our expectation is a phase diagram similar to \fig{fig:summaryrect} for the rectangular case.
We expect a spatial deconfinement transition line with order parameter $P_L$.
\begin{itemize}
  \item In the high-temperature, small-volume limit $r_{\beta}, r_L \ll 1$ we expect $P_L \ne 0$ and thermal behavior as in Eq.~\eqref{eq:HT}.
  \item For high temperatures $r_{\beta}^3 \ll r_L$ the SYM may be dimensionally reduced to the BQM theory, leading us to expect the phase transitions described in Eq.~\eqref{eq:skewBQM}.
        We will have $P_L \ne 0$ for $r_L^3 \lesssim 1.4 r_{\beta} / (1 - \ga^2)^{3 / 2}$, and $P_L = 0$ otherwise.
  \item For low temperatures $t \ll 1$ we expect a IIA or IIB gravity black hole description.
        The D0~phase, with approximate behavior~\eqref{eq:skewD0phase} and $P_L \ne 0$, dominates for $r_L^2 < \cgrav \sqrt{1 - \ga^2} r_{\beta}$.
        For $r_L^2 > \cgrav \sqrt{1 - \ga^2} r_{\beta}$ we expect the D1~phase with $P_L = 0$ to dominate with behavior~\eqref{eq:skewD1phase}, where this formula assumes $r_L$, $r_{\beta}$ and \ga are in the fundamental domain.
\end{itemize}
We expect all these phases to be thermally deconfined, so assuming the parameters $r_L$, $r_{\beta}$ and \ga are in the fundamental domain then we will have $P_{\beta} \ne 0$.
If they are not in the fundamental domain, it is possible in the D1~phase to have $P_{\beta} = 0$ even though $P_{\beta}' \ne 0$ for a fundamental-domain description of the torus, $r_L'$, $r_{\beta}'$ and $\ga'$.

\begin{figure}[tbp]
  \centering
  \includegraphics[height=6 cm]{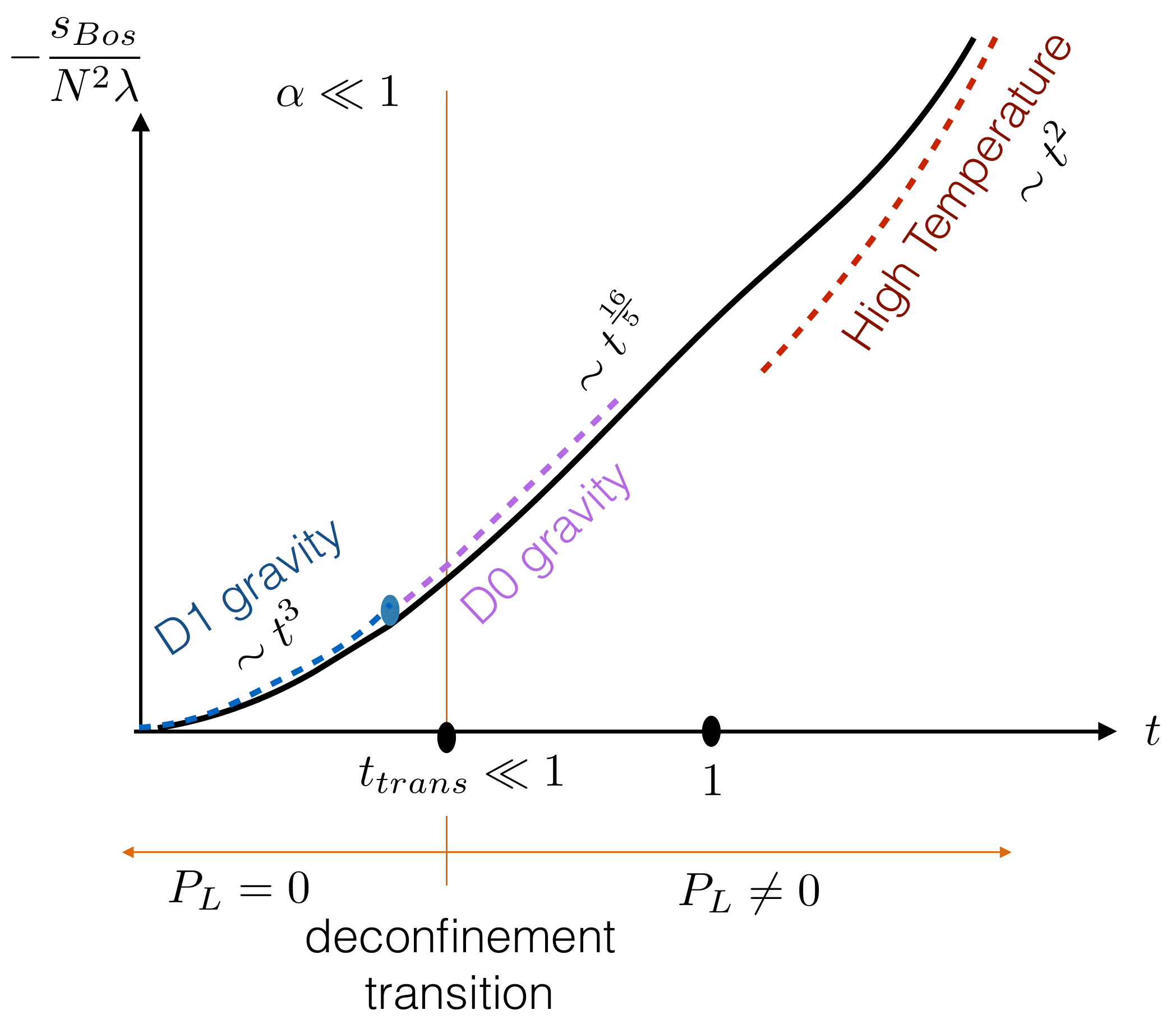}\hfill \includegraphics[height=6 cm]{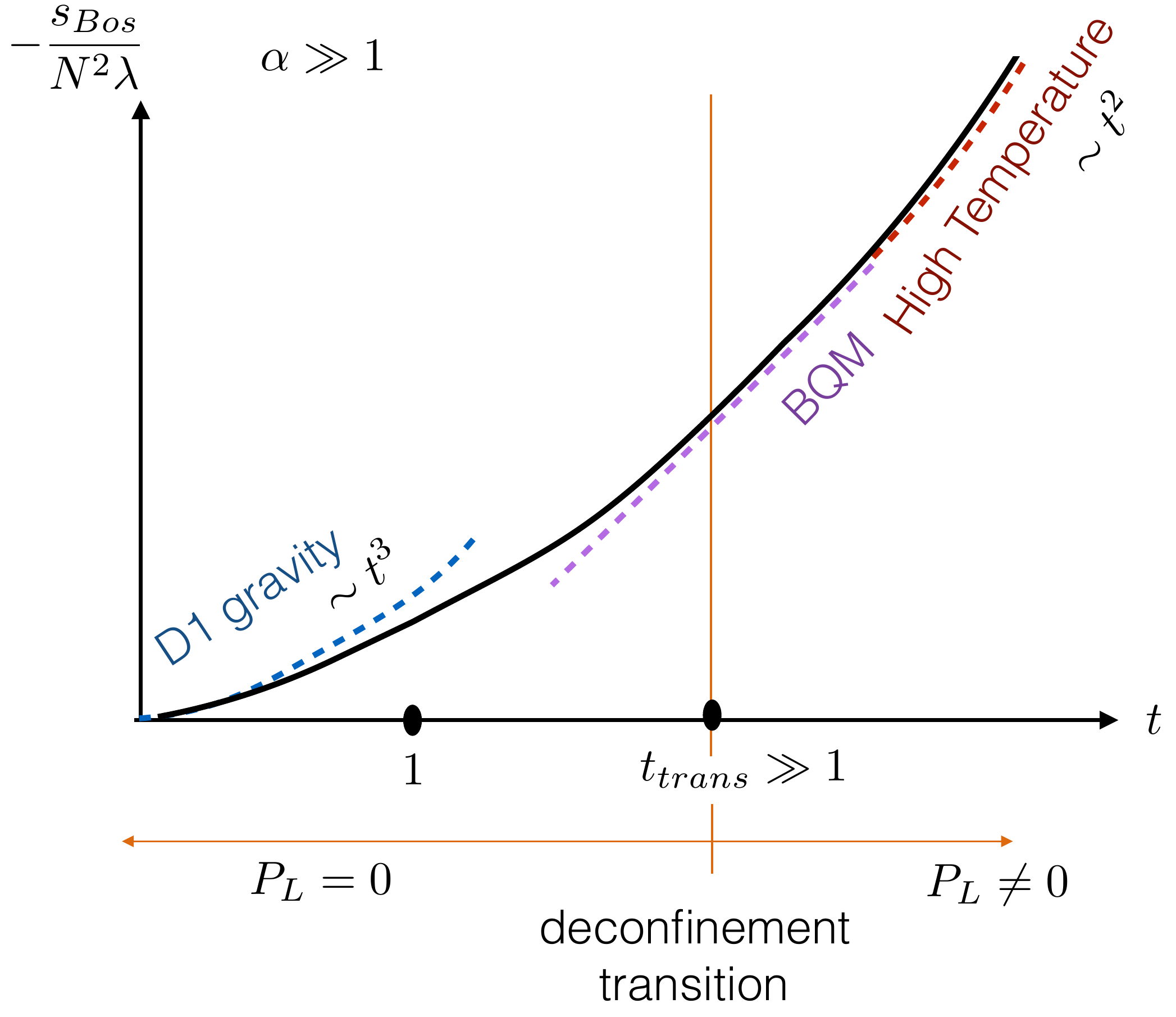} \\[6 pt]
  \includegraphics[height=6 cm]{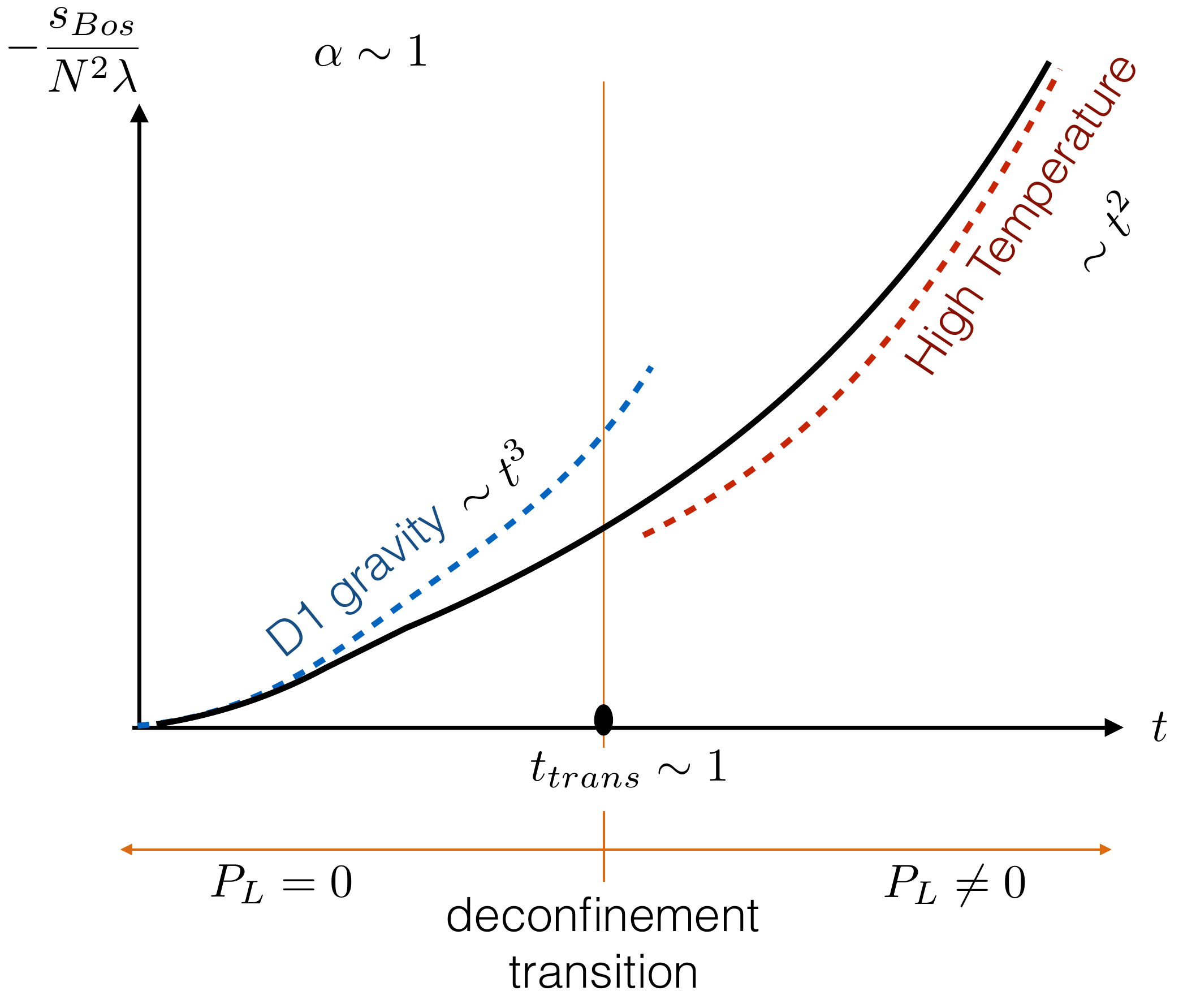}
  \caption{\label{fig:summaryskew}Summary of the expected behavior of SYM on a skewed torus varying $t$ with fixed shape parameters $\al, \ga$.  As $t$ varies from high to low we pass from the small-volume $P_L \ne 0$ deconfined phase into the $P_L = 0$ confined gravity D1~phase.  For small \al (top-left plot) the low-$t$ behavior, including the phase transition, falls in the gravity regime.  Hence we see not only the D1~phase, but also the D0~phase and the first-order transition between them.  For large \al (top-right plot) the high-$t$ behavior including the phase transition to the $P_L = 0$ confined phase is described by the BQM reduction.}
\end{figure}

In our numerical analyses of the skewed SYM theory it will be convenient to fix $\al = r_L / r_{\beta}$ and vary $t = 1 / r_{\beta}$ to scan a `slice' of the $r_{\beta} \times r_L$ plane.
For any finite $\al$, at sufficiently high temperature $t \gg 1$ we will also be in the small-volume regime with $P_L \ne 0$.
As we decrease $t$, for \emph{any} finite \al we expect to go through a confinement phase transition associated to $P_L$, and for $t \ll 1$ eventually enter the gravity D1~phase with $P_L = 0$.
For large $\al \gg 1$ this will be the confinement transition described by BQM.
For small $\al \ll 1$ we expect to enter the gravity regime in the spatially deconfined D0~phase, and encounter the first-order dual Gregory--Laflamme transition to the D1~phase at the lower temperature $t = \al^2 / (\cgrav \sqrt{1 - \ga^2})$.
These expectations for large, small and intermediate \al are illustrated in \fig{fig:summaryskew}.

\section{\label{sec:lattice}Lattice formulation}
Using ideas borrowed from topological field theory and orbifold constructions it has recently become possible to construct a four-dimensional lattice theory which retains an exact supersymmetry at non-zero lattice spacing and produces $\cN = 4$ SYM in the continuum limit.
Noting that maximal SYM in any dimension can be thought of as a classical dimensional reduction of the $\cN = 1$ SYM in 10~dimensions, it follows that our theory of interest, maximal SYM in two dimensions, can be derived from a dimensional reduction of the four-dimensional $\cN = 4$ SYM theory.
Thus the approach we take here is to use the existing four-dimensional lattice construction of $\cN = 4$ SYM, and reduce this in two directions to obtain a two-dimensional lattice action for our desired two-dimensional SYM.

An interesting subtlety arises, namely that the four-dimensional lattice most naturally has an $A_4^*$ geometry rather than a hypercubic one.
When we reduce this lattice action, we obtain a discretization of two-dimensional SYM on a two-dimensional $A_2^*$ (triangular) lattice.
Taking the lattice to be periodic, with one direction having thermal (anti-periodic) fermion BCs and the other periodic BCs, we generate two-dimensional SYM on a 2-torus which is skewed, as the $A_2^*$ lattice basis vectors are not orthogonal.
However, as emphasized, this should be viewed as a virtue rather than a problem.
While there is no direct Lorentzian interpretation of this finite-volume `generalized' thermal ensemble, as discussed above there are holographic string theory predictions that can be tested, and that is the aim of this paper.

We begin by considering the four-dimensional lattice discretization of topologically twisted $\cN = 4$ SYM.
We then outline how this is reduced to a two-dimensional system whose continuum limit will be the reduction of $\cN = 4$ SYM, giving maximal SYM in two dimensions.
Since the reduced lattice has an $A_2^*$ geometry, where we make one lattice direction into the thermal circle of length $\beta$ and the other into the spatial circle of length $L$, the continuum limit will be two-dimensional SYM living on a skewed torus.
The skewing parameter is then determined from the $A_2^*$ lattice geometry to be $\ga = -1 / 2$.
In Appendix~\ref{app:scalar} we provide a detailed discussion of the reduction of a four-dimensional $A_4^*$ lattice theory to the two-dimensional $A_2^*$ lattice for a simpler scalar field theory.

\subsection{\label{sec:lattice_4d}Four-dimensional twisted lattice $\cN = 4$ SYM}
In this section we summarize the important features of this four-dimensional lattice theory before proceeding to its dimensional reduction.
The trick to preserving supercharges in a lattice theory is to discretize a topologically twisted formulation of the underlying supersymmetric theory.\footnote{In the case of gauge theories these lattice formulations were first derived using ideas from orbifolding and deconstruction~\cite{Cohen:2003xe, Cohen:2003qw, Kaplan:2005ta}.}
In the case of $\cN = 4$ SYM the twisted construction treats the four-component gauge field and the six massless adjoint scalars of the theory as a five-component complexified gauge field
\begin{equation}
  \label{complexification}
  \cA_a \equiv A_a + iB_a,
\end{equation}
where the roman index `$a$' runs from 1 to 5.
The four Majorana fermions of the theory are decomposed into $\chi_{ab} = -\chi_{ba}$, $\psi_a$ and $\eta$.
The analogous decomposition of the sixteen supercharges provides a twisted-scalar \cQ corresponding to $\eta$, which is nilpotent, $\cQ^2 = 0$.
The complexified gauge field leads to complexified field strengths
\begin{align}
  \cF_{ab} & \equiv [\cD_a, \cD_b] &
  \cFb_{ab} & \equiv [\cDb_a, \cDb_b],
\end{align}
where the corresponding complexified covariant derivatives are
\begin{align}
  \cD_a & = \partial_a + \cA_a &
  \cDb_a & = \partial_a + \cAb_a.
\end{align}
Using these ingredients we can express the usual $\cN = 4$ action as a sum of $\cQ$-exact and $\cQ$-closed terms,
\begin{equation}
  \label{eq:contSYM}
  \begin{split}
    S & = \frac{N}{4\lam_4} \cQ \int d^4 x \ \Tr{\chi_{ab}\cF_{ab} + \eta [\cDb_a, \cD_a] - \frac{1}{2}\eta d} + S_{\text{cl}} \\
    S_{\text{cl}} & = -\frac{N}{16\lam_4} \int d^4 x \ \Tr{\epsilon_{mnpqr} \chi_{qr} \cDb_p \chi_{mn}},
  \end{split}
\end{equation}
where $\lam_4 = g^2 N$ is the usual 't~Hooft coupling and we implicitly sum over repeated indices.
Here $x_{\al}$ are the usual canonical flat-space coordinates with \al running from 1 to 4.
The action of the scalar supersymmetry charge \cQ is
\begin{align}
  & \cQ\; \cA_a = \psi_a         & & \cQ\; \psi_a = 0 \cr
  & \cQ\; \chi_{ab} = -\cFb_{ab} & & \cQ\; \cAb_a = 0 \\
  & \cQ\; \eta = d               & & \cQ\; d = 0,     \nn
\end{align}
where $d$ is a bosonic auxiliary field with equation of motion $d = \left[\cDb_a, \cD_a\right]$.
Since $\cQ^2 = 0$ the $\cQ$-exact part of the action is clearly supersymmetric, while \cQ acting on the $\cQ$-closed term vanishes due to a Bianchi identity.
The other fifteen supercharges are twisted into a 1-form $\cQ_a$ and antisymmetric 2-form $\cQ_{ab}$.

This continuum action can be discretized while preserving the single \cQ supersymmetry as described in Refs.~\cite{Kaplan:2005ta, Catterall:2007kn, Damgaard:2008pa, Catterall:2014vka}.
This discretization procedure dictates how the continuum fields are placed on the lattice, how derivatives are replaced by lattice difference operators, and even the structure of the underlying lattice itself.
Specifically, we must employ the $A_4^*$ lattice whose five basis vectors symmetrically span the four spacetime dimensions.
This lattice is a natural generalization of the two-dimensional triangular ($A_2^*$) lattice to four dimensions.
It possesses five equivalent basis vectors corresponding to the vectors from the center of an equilateral four-simplex out to its five vertices.
It has a high $S_5$ point group symmetry with the dimensions of its low lying irreducible representations matching those of the continuum twisted SO(4) rotation group.
Ref.~\cite{Catterall:2014mha} shows that the combination of the \cQ supersymmetry, lattice gauge invariance and the $S_5$ global symmetry suffices to ensure that no new relevant operators are generated by quantum corrections.
Assuming non-perturbative effects such as instantons preserve the lattice moduli space, only a single marginal coupling may need to be tuned to obtain $\cN = 4$ SYM in the continuum limit.

The resultant lattice action takes the form
\begin{align}
  S_0 & = \frac{N}{4\lalat} \sum_{\vn} \Tr{\cQ \left(\chi_{ab}(\vn)\cD_a^{(+)}\cU_b(\vn) + \eta(\vn) \cDb_a^{(-)}\cU_a(\vn) - \frac{1}{2}\eta(\vn) d(\vn) \right)} + S_{\text{cl}} \\
  S_{\text{cl}} & = -\frac{N}{16\lalat} \sum_{\vn} \Tr{\epsilon_{abcde} \chi_{de}(\vn + \hatbmu_a + \hatbmu_b + \hatbmu_c) \cDb^{(-)}_{c} \chi_{ab}(\vn + \hatbmu_c)},
\end{align}
where the lattice difference operators appearing in the above expression are given in Refs.~\cite{Catterall:2007kn, Damgaard:2008pa} and generically take the form of shifted commutators.
For example,
\begin{equation}
  \cD_a^{(+)}\cU_b(\vn) = \cU_a(\vn)\cU_b(\vn + a) - \cU_b(\vn)\cU_a(\vn + b).
\end{equation}
Remarkably the $\cQ$-closed term is still lattice supersymmetric due to the existence of an exact lattice Bianchi identity,
\begin{equation}
  \epsilon_{abcde} \cDb^{(-)}_c \cFb_{ab}(\vn + \hat{\mu}_c) = 0.
\end{equation}
Integrating out the auxiliary field $d$ yields
\begin{equation}
  \label{eq:lat_act}
  \begin{split}
    S_0 & = \frac{N}{4\lalat} \sum_{\vn} \text{Tr}\left[-\cFb_{ab}(\vn) \cF_{ab}(\vn) + \frac{1}{2}\left(\cDb_a^{(-)}\cU_a(\vn)\right)^2 - \chi_{ab}(\vn) \cD^{(+)}_{[a}\psi_{b]}(\vn) - \eta(\vn) \cDb^{(-)}_a\psi_a(\vn)\right] + S_{\text{cl}}.
  \end{split}
\end{equation}
The lattice sites in the canonical flat-space coordinates $x^{\al}$ of Eq.~\eqref{eq:contSYM} are arranged as the $A_4^*$ lattice with positions $x^{\al} = \Delta\, \vn^{\nu} e_{(\nu)}^{\al}$ for $\vn \in \Zbb^4$.
This discretization is analogous to that discussed explicitly for the scalar theory example in Appendix~\ref{app:scalar}.
As discussed following Eq.~\eqref{eq:4dact1}, the resulting continuum action~\eqref{eq:contSYM} has a coupling related to the lattice coupling as $\lam_4 = \lalat / \sqrt{5}$~\cite{Kaplan:2005ta, Catterall:2014vka}.

The presence of an exact lattice supersymmetry allows us to derive an exact expression for the renormalized bosonic action density, which gives the derivative of the partition function with respect to the coupling as in Eq.~\eqref{bosonicaction}.
We find
\begin{equation}
  \label{eq:subtraction}
  \vev{s_{\text{Bos}}} = \left( \frac{\vev{S_B^{\text{lat}}}}{V} - \frac{9N^2}{2} \right)
\end{equation}
where $V$ denotes the number of lattice sites and $S_B^{\text{lat}}$ corresponds to the bosonic terms in the lattice action.
This definition of the continuum renormalized $\vev{s_{\text{Bos}}}$ has the property that it vanishes as a consequence of the exact lattice supersymmetry, if periodic (non-thermal) BCs are used.

In practice, to stabilize the SU($N$) flat directions of the theory we add to $S_0$ a soft-supersymmetry-breaking scalar potential
\begin{equation}
  \label{eq:single_trace}
  S_{\text{soft}} = \frac{N}{4\lalat} \mu^2 \sum_{\vn,\ a} \Tr{\bigg(\cUb_a(\vn) \cU_a(\vn) - \Ibb_N\bigg)^2}
\end{equation}
with tunable parameter $\mu$.
In the dimensionally reduced system this term is particularly important at low temperatures where the flat directions lead to thermal instabilities~\cite{Catterall:2009xn}.
This single-trace scalar potential differs from the double-trace operator used in previous investigations~\cite{Catterall:2014vka, Schaich:2014pda, Catterall:2014vga, Catterall:2015ira, Schaich:2015daa, Schaich:2015ppr, Schaich:2016jus} and constrains each eigenvalue of $\cUb_a \cU_a$ individually, rather than only the trace as a whole.
Exact supersymmetry at $\mu = 0$ ensures that all $\cQ$-breaking counterterms vanish as some power of $\mu$.

The complexification of the gauge field in Eq.~\eqref{complexification} leads to an enlarged U($N$) = SU($N$) $\otimes$ U(1) gauge invariance.
In the continuum the U(1) sector decouples from observables in the SU($N$) sector, but this is not automatic at non-zero lattice spacing~\cite{Catterall:2014vka, Schaich:2014pda, Catterall:2014vga}.
To regulate additional flat directions in the U(1) sector, we truncate the theory to remove the U(1) modes from $\cU_a$, making them elements of the group SL($N, \Cbb$) rather than the algebra $\glN$.
In order to maintain SU($N$) gauge invariance it is necessary to keep the fermions in $\glN$, explicitly breaking the lattice supersymmetry that would have related $\cU_a$ to $\psi_a$.
However, by representing the truncated gauge links as $\cU_b = e^{iga \cA_b}$, we can argue that the continuum supersymmetry relating $\cA_a$ and $\psi_a$ is approximately realized in the large-$N$ limit even at non-zero lattice spacing.
This follows from fixing the 't~Hooft coupling $\lalat = g^2 N$ as $N \to \infty$, implying $g^2 \to 0$.
Then expanding the exponential produces the desired $\cU_b = \Ibb_N + iga\cA_b$ up to $\cO(ga)$ corrections that vanish as $N \to \infty$ even at non-zero lattice spacing $a$.
Empirically, when we measure would-be supersymmetric Ward identities we find that they are satisfied up to small (at most percent-level) deviations, and those deviations decrease $\propto 1 / N^2$ as $N$ increases.
(Figure~\ref{fig:ward} in Appendix~\ref{app:num} shows some representative results.)
Although the constant term in Eq.~\eqref{eq:subtraction} is no longer exactly $9 / 2$, we expect only comparably small corrections to it and therefore continue to use Eq.~\eqref{eq:subtraction} to define $s_{\text{Bos}}$.
Since our lower-dimensional studies all focus on holographic dualities in the large-$N$ limit, this truncated approach appears viable at least in fewer than four dimensions.

\subsection{\label{sec:lattice_2d}Two-dimensional twisted lattice $\cN = (8, 8)$ SYM}
Our interest is in two-dimensional maximal SYM on a 2-torus with thermal (anti-periodic) fermion BCs on one cycle and periodic BCs on the other.
This two-dimensional maximal SYM is given by the dimensional reduction of the four-dimensional $\cN = 4$ theory.
Hence to obtain this two-dimensional theory on a 2-torus we simply consider the above four-dimensional lattice discretization of the $\cN = 4$ theory, taken on $N_x \times 1 \times 1 \times N_t$ lattices with periodic BCs in the reduced directions, corresponding to naive dimensional reduction.
The gauge fields associated to the reduced directions now transform as site fields $\varphi_i(\vn)$ and are naturally interpreted as the scalar fields arising from dimensional reduction.
Their fermionic superpartners, now also site fields, correspond to additional exact lattice supersymmetries~\cite{Damgaard:2008pa}.

Exactly as for the scalar example discussed in Appendix~\ref{app:scalar}, such a reduction results in a continuum theory in two-dimensional flat space where the lattice geometry is that of $A_2^*$.
We take appropriate periodicity conditions on the extended lattice directions, $\vn \sim \vn + (0, 0, 0, N_t)$ with anti-periodic fermions and $\vn \sim \vn + (N_x, 0, 0, 0)$ with periodic fermions, thus generating the 2-torus.
Since the two-dimensional lattice has $A_2^*$ geometry the 2-torus we generate is not rectangular but skewed, with $\ga = \frac{\tilde{\ve}_1 \cdot \tilde{\ve}_2}{|\tilde{\ve}_1| |\tilde{\ve}_2|} = -1 / 2$.
The lengths of the two cycles are then
\begin{align}
  \beta & = a_2 \, N_t &
  L & = a_2 \, N_x,
\end{align}
where we see from Appendix~\ref{app:scalar} that the two-dimensional lattice spacing is $a_2 = \Delta \sqrt{2 / 3}$.
Similarly, the two-dimensional continuum gauge coupling is
\begin{equation}
  \lam = \frac{\lalat}{\Delta^2 \sqrt{3}} = \frac{\lam_4}{\Delta^2} \sqrt{\frac{5}{3}},
\end{equation}
where the second equality from Eq.~\eqref{eq:KKcoupling} considers the coupling as arising from an appropriate Kaluza--Klein reduction of the continuum four-dimensional $\cN = 4$ theory.
Thus in terms of our dimensionless couplings our lattice action corresponds to two-dimensional $\cN = (8, 8)$ SYM on a skewed torus with $\ga = -1 / 2$ and
\begin{align}
  r_{\beta} & = \beta \sqrt{\lam} = N_t \sqrt{\frac{2\lalat}{3\sqrt{3}}} &
  r_L & = L \sqrt{\lam} = N_x \sqrt{\frac{2\lalat}{3\sqrt{3}}},
\end{align}
which, being dimensionless, are independent of the scale $\Delta$ as they should be.
Noting that in two dimensions the continuum SYM is super-renormalizable, we do not expect any renormalization of the classical geometry of this 2-torus.

Finally, we add an additional soft-$\cQ$-breaking term to ensure the dimensionally reduced lattice theory correctly reproduces the physics of the continuum theory:
\begin{equation}
  \label{eq:center}
  S_{\text{center}} = -\frac{N}{4\lalat} c_W^2 \sum_{\vn,\ i = y, z} 2\text{ReTr}\bigg[\varphi_i(\vn) + \varphi_i^{-1}(\vn)\bigg].
\end{equation}
This is gauge invariant since $\varphi_i(\vn)$ transform as site fields.
In the absence of this term we observe correlated instabilities in the scalar eigenvalues and in $\Tr{\varphi_i}$ for low dimensionless temperatures $t \lesssim 1$.
This is not the correct behavior required by Kaluza--Klein reduction in the continuum.
Instead it corresponds to a center-symmetric phase for the reduced dimensions, which could correspond to Eguchi--Kawai reduction at large~$N$ in the presence of adjoint fermions.
We avoid this center-symmetric phase by using $S_{\text{center}}$ to explicitly break the center symmetry.
In this work we use $c_W^2 = \mu^2$ for low $t \lesssim 1$ and $c_W^2 = 0$ for high temperatures $t \gg 1$, again extrapolating $\mu^2 \to 0$ in the former case.

\subsection{\label{sec:torusgeom}Torus geometries}
The geometry of our tori is determined by the skewing parameter $\ga = -1 / 2$ set by our lattice discretization, and by the aspect ratio $\al = r_L / r_{\beta} = N_x / N_t$, where $N_x$ and $N_t$ are respectively the numbers of lattice points generating the spatial and temporal cycles.
As mentioned earlier, we will typically discuss results specifying the torus with the skewing $\ga = -1 / 2$, but this parameterization may represent a modular parameter outside the fundamental domain.
Here we review the geometries we will consider and their fundamental parameterization.
In particular, while we use a skewed lattice, some of our geometries in fact are those of rectangular tori when mapped to the fundamental domain.

\begin{table}[htbp]
  \centering
  \renewcommand\arraystretch{1.2}   
  \addtolength{\tabcolsep}{1 pt}    
  \begin{tabular}{c|c|c|c|c|c}
    \al     & $N_x \times N_t$            & Modular Transformation                                            & $\left(\al', \ga'\right)$                                & $t' / t$             &             \\
    \hline
    $1 / 2$ &  $6\times 12$, $8\times 16$ & $\vec{\beta}' = \vec{\beta} + \vec{L}$, $\vec{L}' = \vec{L}$      & $\left(\frac{1}{\sqrt{3}}, 0\right)$                     & $\frac{2}{\sqrt{3}}$ & Rectangular \\
    1       &  $8\times 8$, $16\times 16$ & ---                                                               & ---                                                      & ---                  & Skewed      \\
    $3 / 2$ & $12\times 8$, $18\times 12$ & ---                                                               & ---                                                      & ---                  & Skewed      \\
    2       & $16\times 8$, $24\times 12$ & ---                                                               & ---                                                      & ---                  & Skewed      \\
    $8 / 3$ & $16\times 6$, $24\times 9$  & $\vec{L}' = \vec{L} + 2\vec{\beta}$, $\vec{\beta}' = \vec{\beta}$ & $\left(\frac{2\sqrt{13}}{3}, \frac{1}{\sqrt{13}}\right)$ & 1                    & Skewed      \\
    4       & $16\times 4$, $24\times 6$  & as above                                                          & $\left(2\sqrt{3}, 0\right)$                              & 1                    & Rectangular \\
    6       & $24\times 4$                & as above                                                          & $\left(2\sqrt{7}, -\frac{1}{2\sqrt{7}}\right)$           & 1                    & Skewed      \\
    8       & $32\times 4$                & $\vec{L}' = \vec{L} + 4\vec{\beta}$, $\vec{\beta}' = \vec{\beta}$ & $\left(4\sqrt{3}, 0\right)$                              & 1                    & Rectangular \\
  \end{tabular}
  \caption{\label{tab:tori}The lattice geometries we numerically analyze.  Our lattice discretization naturally picks $\ga = -1 / 2$, and by varying the spatial and temporal lattice extents $N_x$ and $N_t$ we generate tori with different aspect ratios $\al$.  When these $(\al, \ga)$ denote a torus with modular parameter $\uptau$ outside the fundamental domain, we give an appropriate modular transformation (as in Eq.~\eqref{eq:SLtransform}) so that the equivalent $(\al', \ga')$ lie within it.  We also give the relation between the temperatures $t' / t$.  The last column states whether the torus, viewed from the fundamental domain, is skewed or rectangular.}
\end{table}

\begin{figure}[tbp]
  \centering
  \includegraphics[height=\figheight]{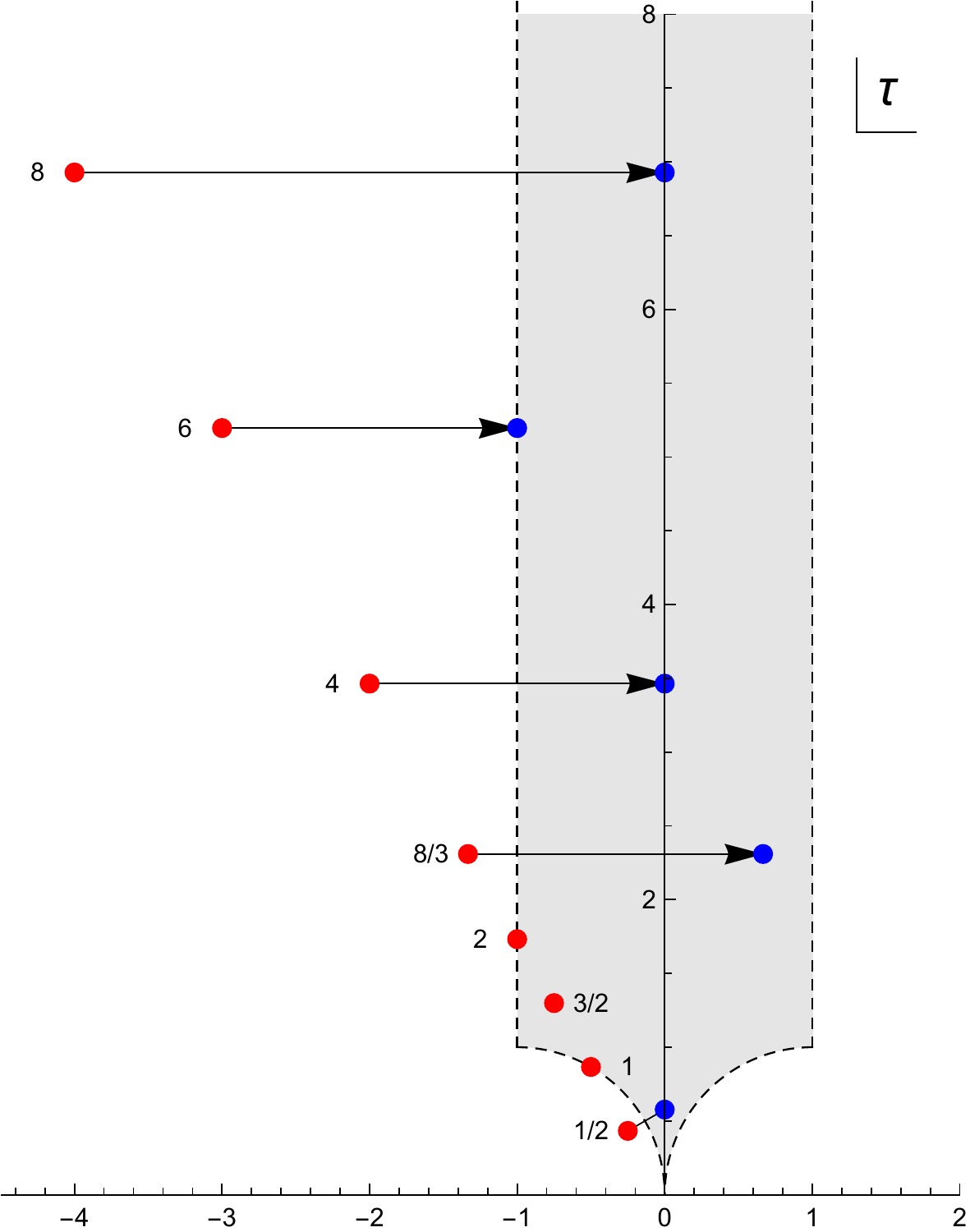}
  \caption{\label{fig:tauplot}Plot of the torus modular parameters $\uptau$ in the complex plane for the aspect ratios \al we numerically analyze.  The red points are for the $\ga = -1 / 2$ of our lattice discretization, with the corresponding \al written next to them.  The fundamental domain is shaded, and when a point lies outside it the equivalent $\uptau'$ lying within it is shown as a blue point.}
\end{figure}

In \tab{tab:tori} we list the lattice sizes $N_x \times N_t$ we numerically analyze, together with their shape parameter \al for skewing $\ga = -1 / 2$.
When the corresponding modular parameter $\uptau$ doesn't fall in the fundamental domain we give a modular transformation to an equivalent representation with shape parameters $\al', \ga'$, and note whether the fundamental representation is rectangular or skewed.
We also give $t' / t$, the ratio between the dimensionless temperature in the new representation to that of the original.
In the corresponding \fig{fig:tauplot} we plot the complex $\uptau$ parameters for the various tori in the natural representation where $\ga = -1 / 2$, and in the cases where these lie outside the fundamental domain we draw an equivalent $\uptau'$ contained in it.

\section{\label{sec:results}Numerical results}
We now discuss our numerical results obtained using the lattice formulation described above.
Before studying the low-temperature regime relevant for supergravity, we first consider the high-temperature, small-volume limit and then the phase structure of the theory.

\subsection{High-temperature limit}
Fixing the shape of the torus, with constant \al for $\ga = -1 / 2$, we vary $t \to \infty$.
Following our earlier discussion in \secref{sec:summaryskew}, this is the high-temperature, small-volume limit where we expect the theory to be spatially deconfined with $P_{\beta}, P_L \ne 0$, and to have bosonic action density~\eqref{eq:HT}.

\begin{figure}[tbp]
  \centering
  \includegraphics[height=\figheight]{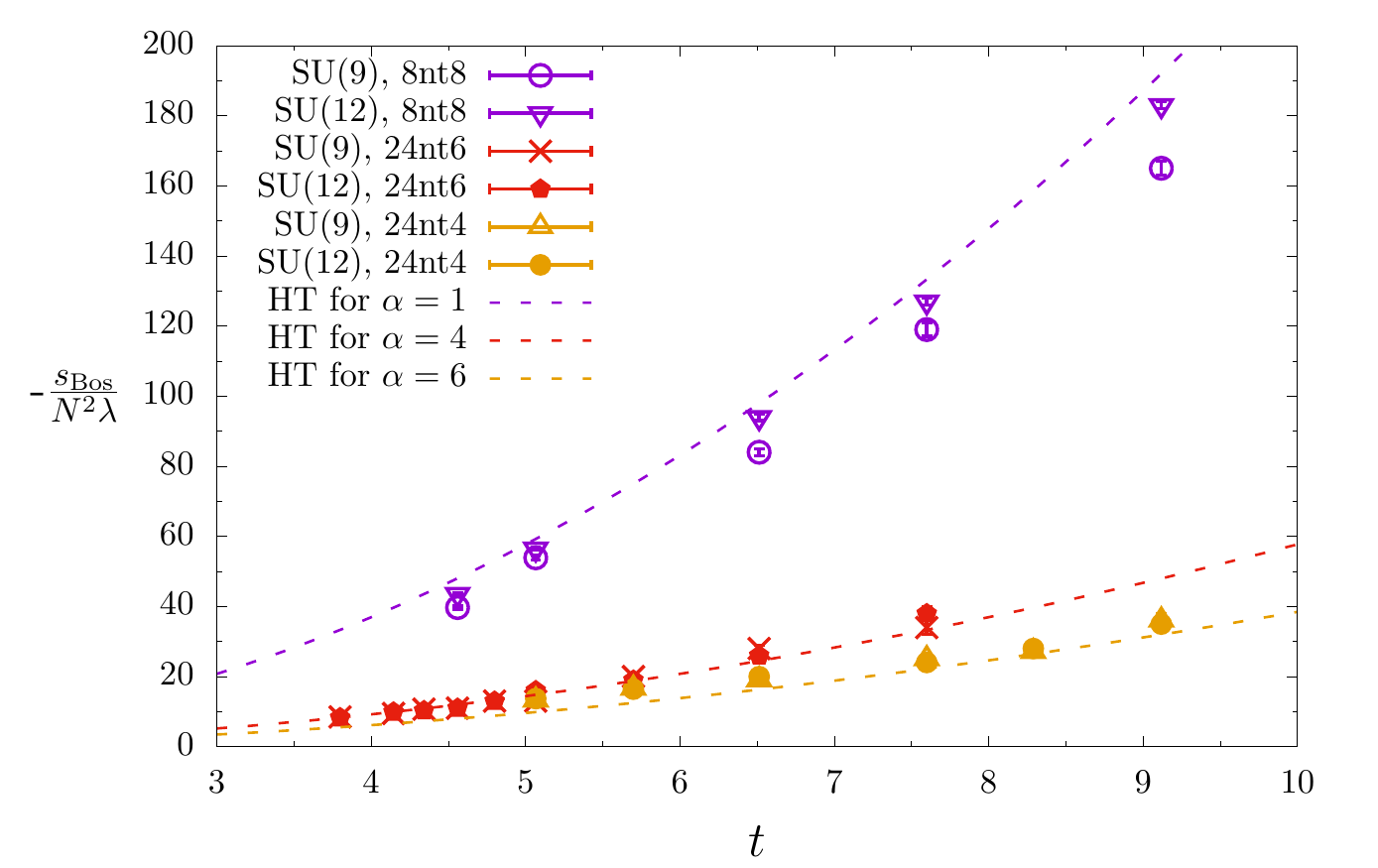}
  \caption{\label{fig:highT}Bosonic action density versus dimensionless temperature $t$ for three aspect ratios $\al = 1$, 4 and 6 (from top to bottom), considering gauge groups SU(9) and SU(12).  The temperature range probed here corresponds to the high-temperature, small-volume limit, and the prediction~\protect\eqref{eq:HT} for the behavior is given by the dashed curves marked HT.}
\end{figure}

We investigate three different aspect ratios ($\al = 1$, 4 and 6) in the high-temperature regime and plot the bosonic action density in \fig{fig:highT}.
Qualitative agreement is seen both in the power of $t$ and the $\al$-dependent coefficient, providing a test of the dimensional reduction that relates the lattice coupling \lalat to the dimensionless continuum parameters $r_L$ and $r_{\beta}$.

\begin{figure}[tbp]
  \centering
  \includegraphics[height=\figheight]{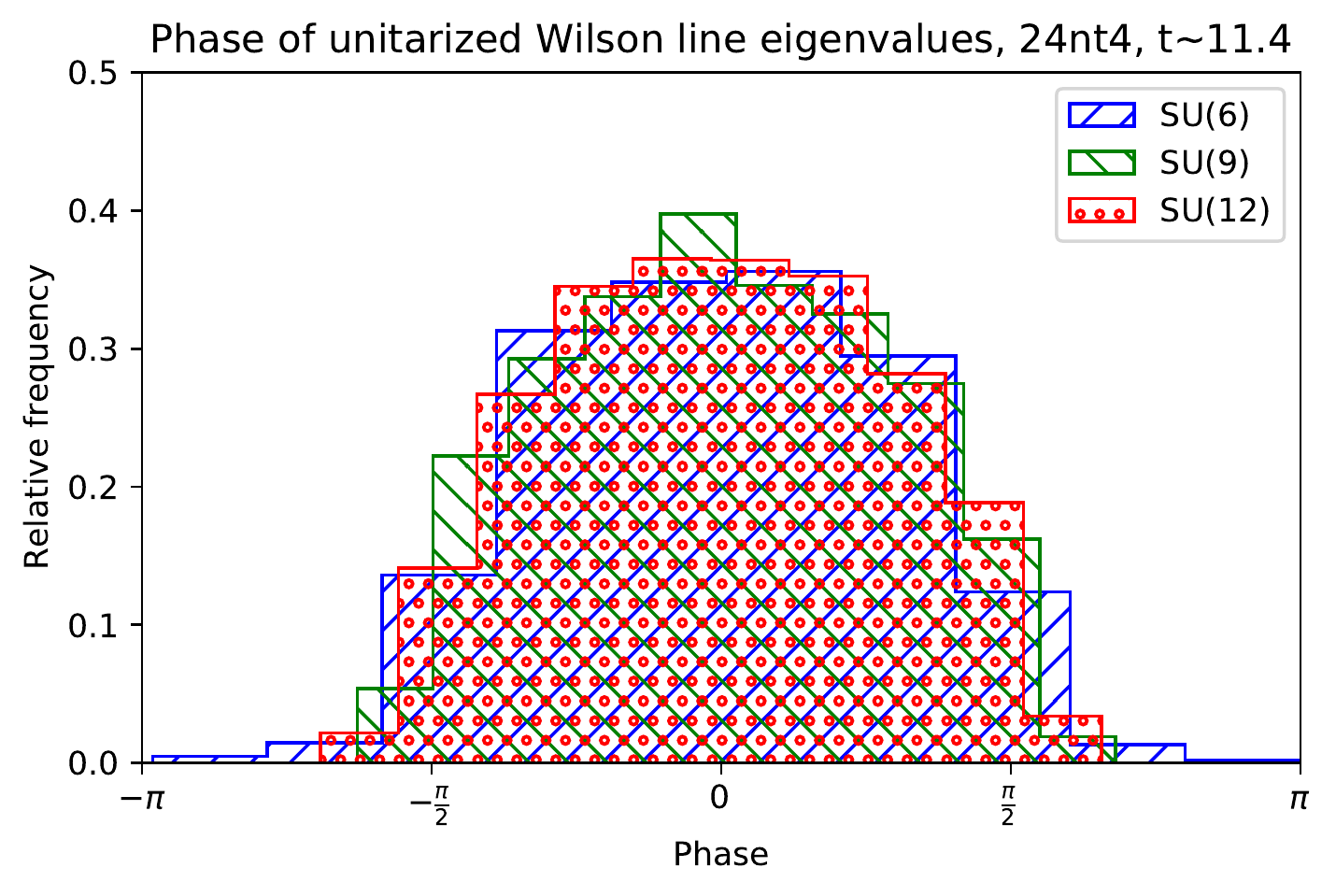}
  \caption{\label{fig:WLeig-highT}Distributions of the phases of the $N$ eigenvalues of spatial Wilson line on $24\times 4$ lattices at a high temperature $t \approx 11.4$, for SU($N$) gauge groups with $N = 6$, 9 and 12.  The phases are measured relative to the average phase of each Wilson line.  The compact distributions correspond to broken $Z_N$ center symmetry in the spatially deconfined high-temperature phase.}
\end{figure}

In \fig{fig:WLeig-highT} we show distributions of the phases of the $N$ eigenvalues of spatial Wilson lines $\cP e^{i\oint_L A}$ on $24\times 4$ lattices ($\al = 6$) at a high temperature $t \approx 11.4$, for SU($N$) gauge groups with $N = 6$, 9 and 12.
The phases are measured relative to the average phase of each Wilson line.
In order to compute the usual Wilson lines from the complexified gauge links $\cU_a$ of the lattice formulation, we use a polar decomposition $\cU_a = H_a \cdot U_a$ to separate each link into a positive-semidefinite hermitian matrix $H_a$ (containing the scalar fields) and a unitary matrix $U_a$ corresponding to the gauge field.
To compute the Wilson lines we simply multiply the unitary matrices, $\prod_{i = 1}^{N_x} U_x(x_i, \tau)$ and $\prod_{i = 1}^{N_t} U_t(x, \tau_i)$.
We construct $P_L$ and $P_{\beta}$ by taking the trace (normalized to 1), averaging over lattice sites in the temporal and spatial direction (respectively), and then computing the ensemble average of the magnitude.
The expectation that $P_L \sim 1$ implies we should expect a localized distribution of the phases of the spatial Wilson line eigenvalues, which is consistent with the results in \fig{fig:WLeig-highT}.
The distributions show little dependence on $N$, though the $N = 6$ case has a few outliers with large fluctuations from the average phase.
As $t$ decreases we expect a transition with $P_L \to 0$, with the eigenvalue distribution spreading over the angular circle and becoming uniform on it.
For $t \lesssim 9$ we do indeed see the distributions spread out over the full angular period, as we discuss in more detail below. 

\subsection{Phase structure of the SYM theory}
\begin{figure}[tbp]
  \centering
  \includegraphics[height=\figheight]{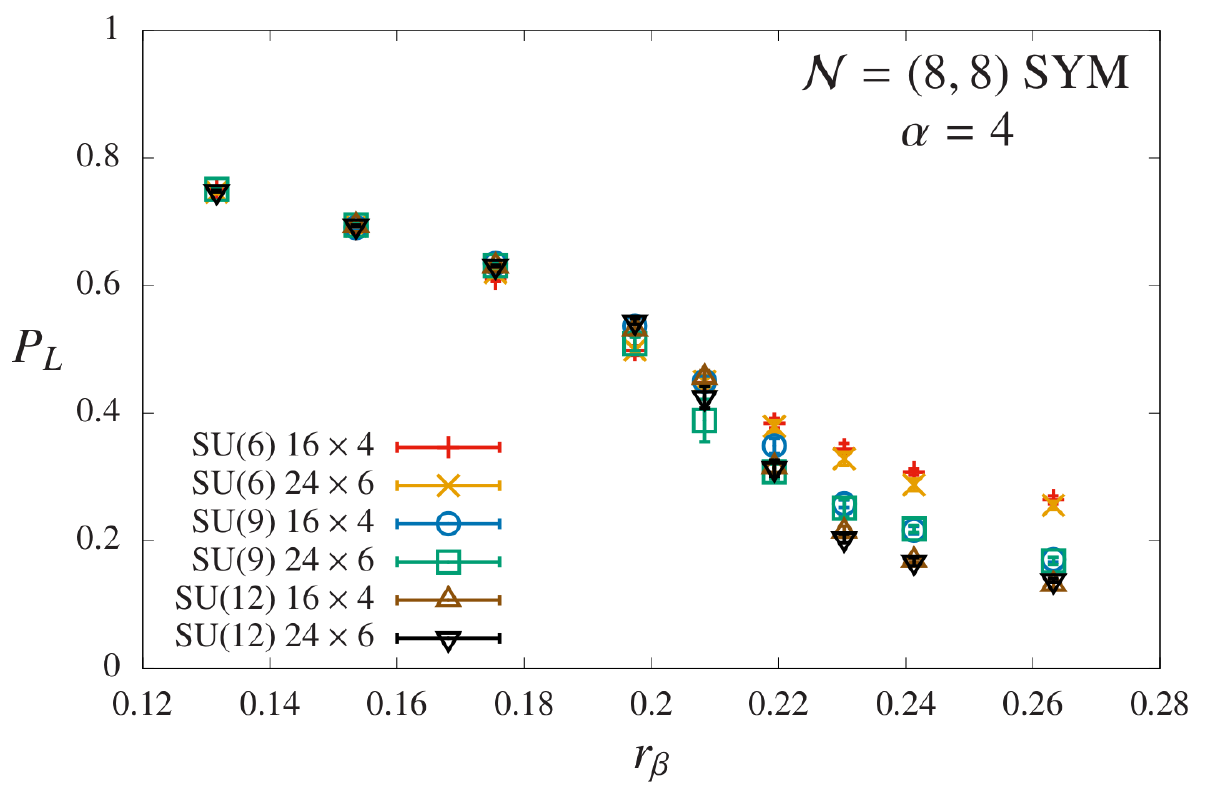}\hfill \includegraphics[height=\figheight]{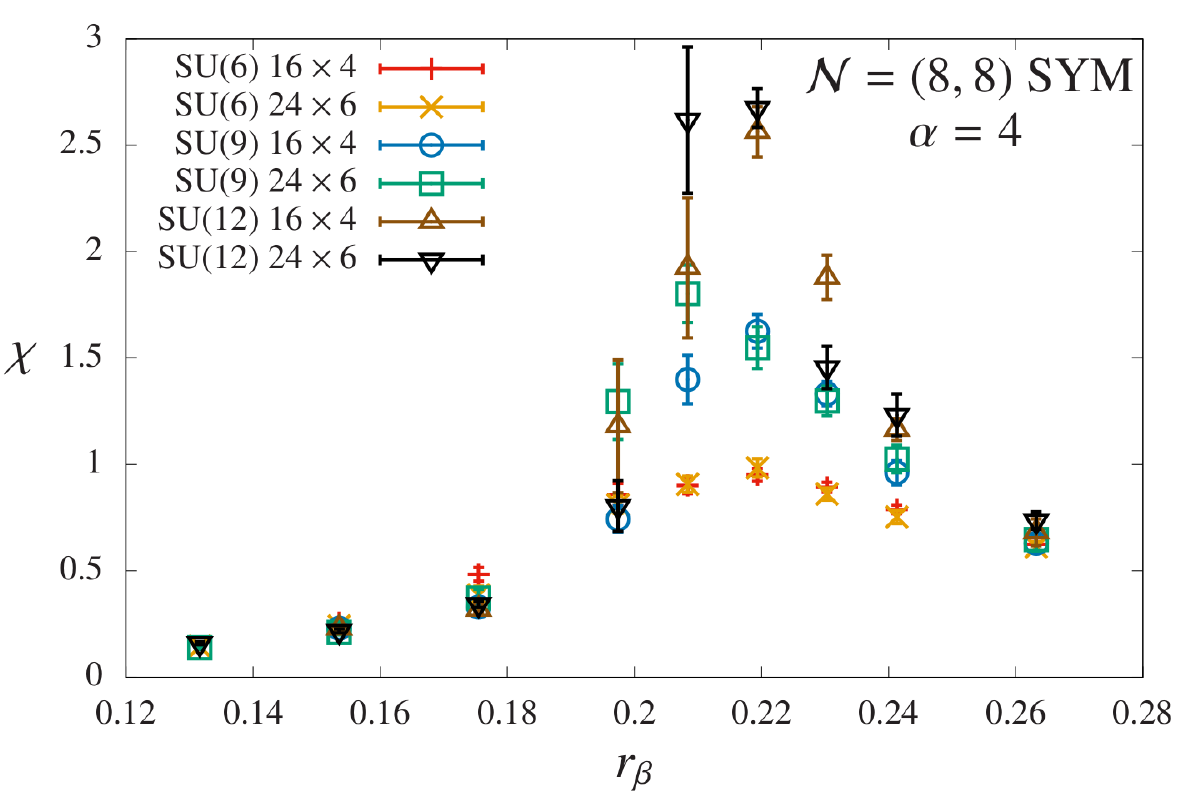}
  \caption{\label{fig:lines_sus_alpha4}Spatial Wilson loop magnitude (left) and susceptibility (right) vs.~inverse dimensionless temperature $r_{\beta} = 1 / t$ for SU($N$) gauge groups with $N = 6$, 9 and 12 on $16\times 4$ and $24\times 6$ lattices (aspect ratio $\al = 4$).  The transition strengthens as $N$ increases, while showing little sensitivity to the lattice size.}
\end{figure}

We have explored the phase structure of the SYM theory by scanning in $t = 1 / r_{\beta}$ for fixed aspect ratio $\al = r_L / r_{\beta}$.
From our previous discussion we expect the theory to be thermally deconfined, but to have an interesting phase structure associated with spatial confinement.
Our numerical results for the temporal Wilson loop magnitude $P_{\beta}$ appear consistent with the theory being thermally deconfined.
We now focus on the spatial Wilson line and order parameter $P_L$.

In \fig{fig:lines_sus_alpha4} we show the jackknife average magnitude of the Wilson line $P_L$ vs.~$r_{\beta}$ for $\al = 4$, along with the corresponding susceptibility
\begin{equation}
  \chi = \vev{\left| \Tr{\cP e^{i\oint_L A}} \right|^2} - \vev{\left| \Tr{\cP e^{i\oint_L A}} \right|}^2.
\end{equation}
The results indicate a large-$N$ transition at $t_c = 4.6(2)$ separating a spatially deconfined phase with $P_L \ne 0$ at small $r_{\beta}$ (high temperatures) from a spatially confined phase at large $r_{\beta}$ (low temperatures) where $P_L \to 0$ as $N \to \infty$.
This transition strengthens with larger $N$, while the general agreement between results from $16\times 4$ and $24\times 6$ lattices indicates that discretization effects are small.
As discussed in \secref{sec:torusgeom} (\tab{tab:tori}), the geometry $\al = 4$ for $\ga = -1 / 2$ is equivalent to a rectangular ($\ga' = 0$) torus with $\al' = 2\sqrt{3}$ and $r_{\beta}' = r_{\beta}$.

\begin{figure}[tbp]
  \centering
  \includegraphics[height=\figheight]{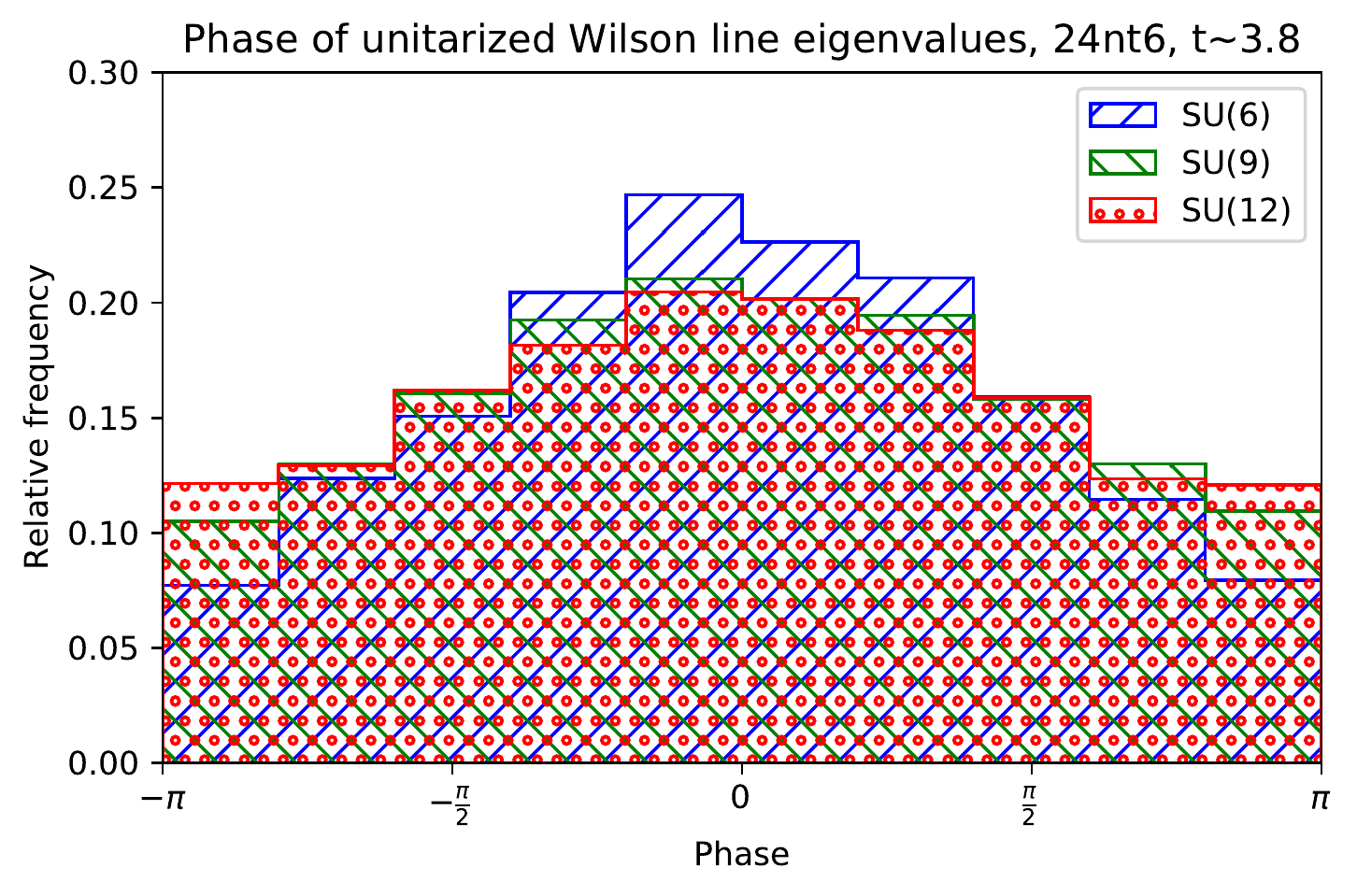}
  \caption{\label{fig:WLeig-lowT}Distributions of Wilson line eigenvalue phases, as in \protect{\fig{fig:WLeig-highT}}, for $24\times 6$ lattices at a lower temperature $t \approx 3.8$.  The distributions are no longer compact, and instead spread throughout the angular period, as expected from the black-string phase of the gravity dual.}
\end{figure}

In \fig{fig:WLeig-lowT} we plot distributions of the spatial Wilson line eigenvalue phases, following the same procedure as described for \fig{fig:WLeig-highT} while considering a lower temperature $t \approx 3.8$ on $24\times 6$ lattices ($\al = 4$).
Since $t < t_c$ we expect to be in a spatially confined phase, with $P_L \to 0$ and correspondingly a uniform density of eigenvalue phases on the angular circle.
As expected, we do observe these phases spreading out around the angular period in \fig{fig:WLeig-lowT}, and the distribution becomes more uniform as $N$ increases.
This contrasts with the localized distributions in \fig{fig:WLeig-highT} for the high-temperature spatially deconfined phase.

\begin{figure}[tbp]
  \centering
  \includegraphics[height=\figheight]{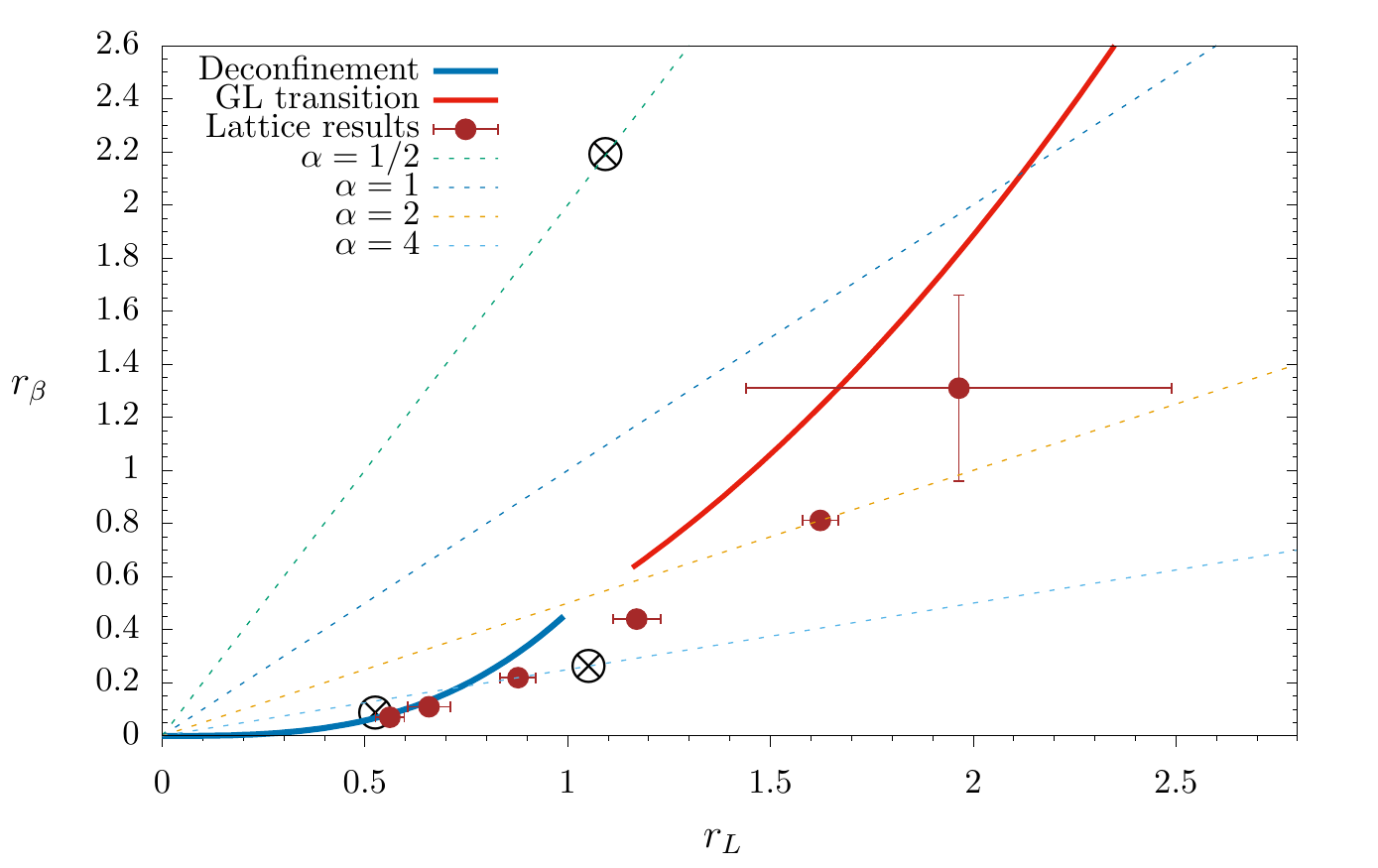}
  \caption{\label{fig:crit}The points show the $(r_L, r_{\beta})$ positions of the SU(12) spatial deconfinement transitions for six aspect ratios $8 \geq \al \geq 3 / 2$ (from left to right), determined from our lattice calculations of the Wilson line susceptibility (cf.\ \protect\fig{fig:lines_sus_alpha4}).  The three $\bigotimes$ symbols mark the ensembles whose Wilson line eigenvalue phase distributions we show in Figs.~\protect\ref{fig:WLeig-highT}, \protect\ref{fig:WLeig-lowT} and \protect\ref{fig:WLeig-al05} (from bottom to top; the point for \protect\fig{fig:WLeig-al2} lies outside the range of the plot).  The solid lines show the expected transitions sketched in \protect\fig{fig:summaryrect}: the BQM deconfinement transition at high temperature and the low-temperature Gregory--Laflamme transition.  The dashed lines indicate constant aspect ratios $1 / 2 \leq \al \leq 4$ from top to bottom.}
\end{figure}

Using the Wilson line susceptibility we have mapped the position of the spatial deconfinement phase transition as a function of \al for our $\ga = -1 / 2$.
In \fig{fig:crit} we plot our results on the $r_L$--$r_{\beta}$ plane and compare them with the expected transitions sketched in \fig{fig:summaryrect}.
The error bars in this figure are the full widths at half maximum of the susceptibility peaks like those shown in the right panel of \fig{fig:lines_sus_alpha4}, considering gauge group SU(12) and a fixed lattice size for each aspect ratio $\al$.
We have checked that alternate determinations of the transition produce consistent results.
These include identifying the transition as the $r_{\beta}$ for which $P_L = 0.5$, motivated by \refcite{Aharony:2004ig}, and using a large-$N$ generalization~\cite{Hudspith:2017BU} of the separatrix introduced for SU(3) by \refcite{Francis:2015lha}.

For $\al \gtrsim 4$ we find the transition occurs at high temperatures $r_{\beta} \ll 1$, and nicely agrees with the deconfinement transition behavior predicted by the high-temperature BQM limit we discussed in \secref{sec:smallcircle}.
Unfortunately, the error bars increase significantly as we approach transitions occurring at lower temperatures (and smaller $\al$) where we expect the dual gravitational prediction~\eqref{eq:skewtransition} to apply.
We do not obtain usable susceptibility peaks for $\al \leq 1$ and cannot conclusively determine the order of the transition for any $\al$.
Given their large uncertainties, our results are certainly consistent with the low-temperature behavior predicted by holography, though we are not yet able to test this prediction with great accuracy.
There are also systematic uncertainties to be considered.
We expect the most significant systematic effects to be those arising from working with gauge group SU(12) and a fixed lattice size for each aspect ratio \al rather than extrapolating to the large-$N$ continuum limit.
While the mild discretization artifacts and $N$ dependence shown in \fig{fig:lines_sus_alpha4} give us confidence that these systematic effects will not qualitatively change our results, we are not yet in a position to meaningfully quantify them.
Carrying out controlled extrapolations to the limits of $N \to \infty$ and infinitely large lattices is a central goal for future generations of lattice calculations.

To summarize, our numerical results for the phase diagram of the two-dimensional SYM system are consistent with the expectations from holography.
We see a phase where the eigenvalues of the spatial Wilson line are uniformly distributed around the unit circle, as expected for a spatially confined phase.
This is separated by a transition from a spatially deconfined phase with localized eigenvalue distribution.
We now study the thermodynamics of the system at low temperatures in both of these phases, comparing to the predictions from the dual gravity theory.

\subsection{D1-phase thermodynamics}
\begin{figure}[tbp]
  \centering
  \includegraphics[height=\figheight]{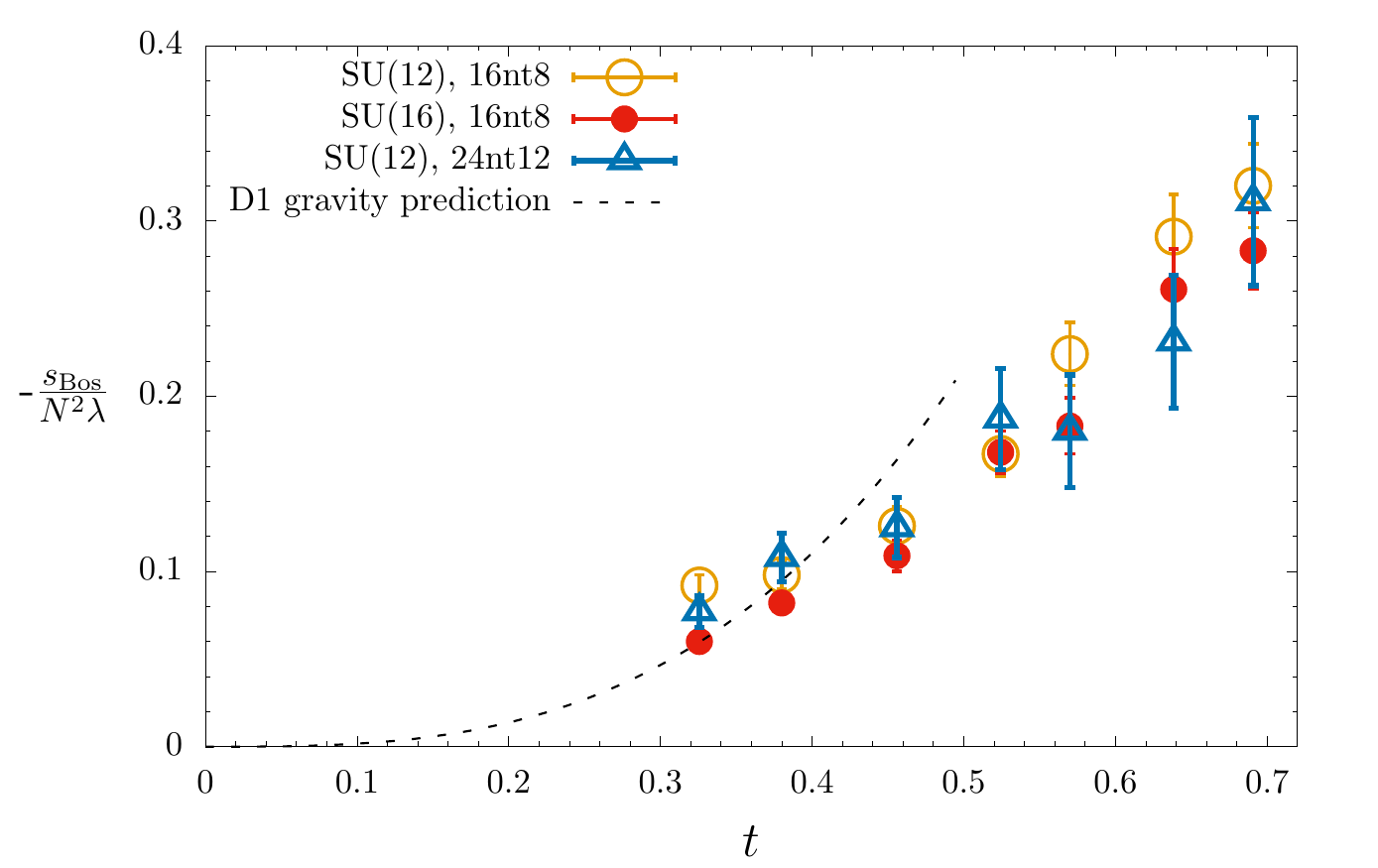}
  \caption{\label{fig:alpha2}Bosonic action density versus dimensionless temperature $t$ for $16\times 8$ and $24\times 12$ lattices (aspect ratio $\al = 2$) with gauge groups SU(12) and SU(16).  All points are results of $\mu^2 \to 0$ extrapolations.  For sufficiently small $t$ our results are consistent with the D1-phase gravity prediction, without significant sensitivity to $N$ or the lattice size.}
\end{figure}
\begin{figure}[tbp]
  \centering
  \includegraphics[height=\figheight]{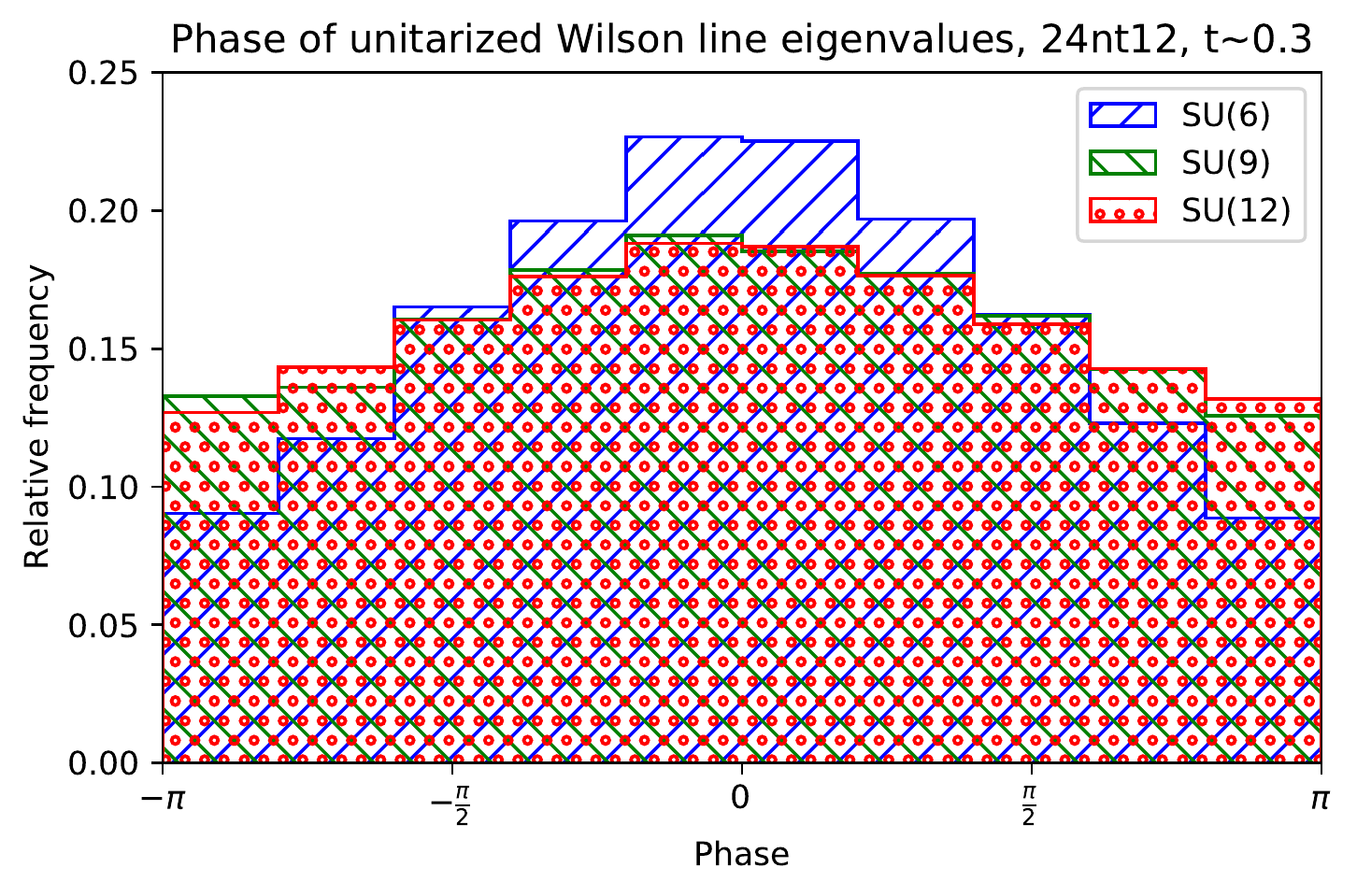}
  \caption{\label{fig:WLeig-al2}Distributions of Wilson line eigenvalue phases, as in \protect{\fig{fig:WLeig-highT}}, for $24\times 12$ lattices at $t \approx 0.33$ with $\mu^2 \approx 0.007$.  The extended distributions, which become more uniform as $N$ increases, correspond to the D1~phase of the gravity dual.}
\end{figure}

In \fig{fig:alpha2} we show the bosonic action density versus $t$ for $\al = 2$ lattice sizes $16\times 8$ and $24\times 12$ with gauge groups SU(12) and SU(16).
From the discussion above, the temperature range shown in the figure lies below the spatial deconfinement transition temperature $t_c \simeq 1.2$.
Hence we expect the system should be spatially confined.
In the low-temperature limit $t \to 0$, we expect it to be described by the gravity D1~phase given in Eq.~\eqref{eq:skewD1phase}.\footnote{Eq.~\eqref{eq:skewD1phase} applies directly since $\al, \ga$ lie in the fundamental domain.}
This is corroborated by our lattice results, which lie close to the D1-gravity prediction at low temperatures, $t \lesssim 0.4$.

In a manner similar to the well-studied case of supersymmetric quantum mechanics, this system becomes unstable for very low temperatures, and hence our calculations do not extend all the way down to $t \lesssim 1 / N$.
The origin of this instability is well understood and is discussed in \refcite{Catterall:2009xn}.
The scalar potential terms~\eqref{eq:single_trace} and \eqref{eq:center} help stabilize our numerical calculations, but still do not allow access to arbitrarily low temperatures.
We then need to extrapolate $\mu^2 \to 0$ to remove the soft-$\cQ$-breaking effects of these terms (with $c_W^2 = \mu^2$), which produces the results shown in \fig{fig:alpha2}.
Simple linear $\mu^2 \to 0$ extrapolations generally have good quality with small $\chidof$; we include a representative example in Appendix~\ref{app:num} (\fig{fig:extrap}).
Figure~\ref{fig:WLeig-al2} shows extended distributions for the Wilson line eigenvalue phases on $24\times 12$ lattices at $t \approx 0.33$ (with $\mu^2 \approx 0.007$), which become more uniform as $N$ increases.
This behavior supports our conclusion that the system is spatially confined in this region of the phase diagram, corresponding to the gravity D1~phase.

\begin{figure}[tbp]
  \centering
  \includegraphics[height=\figheight]{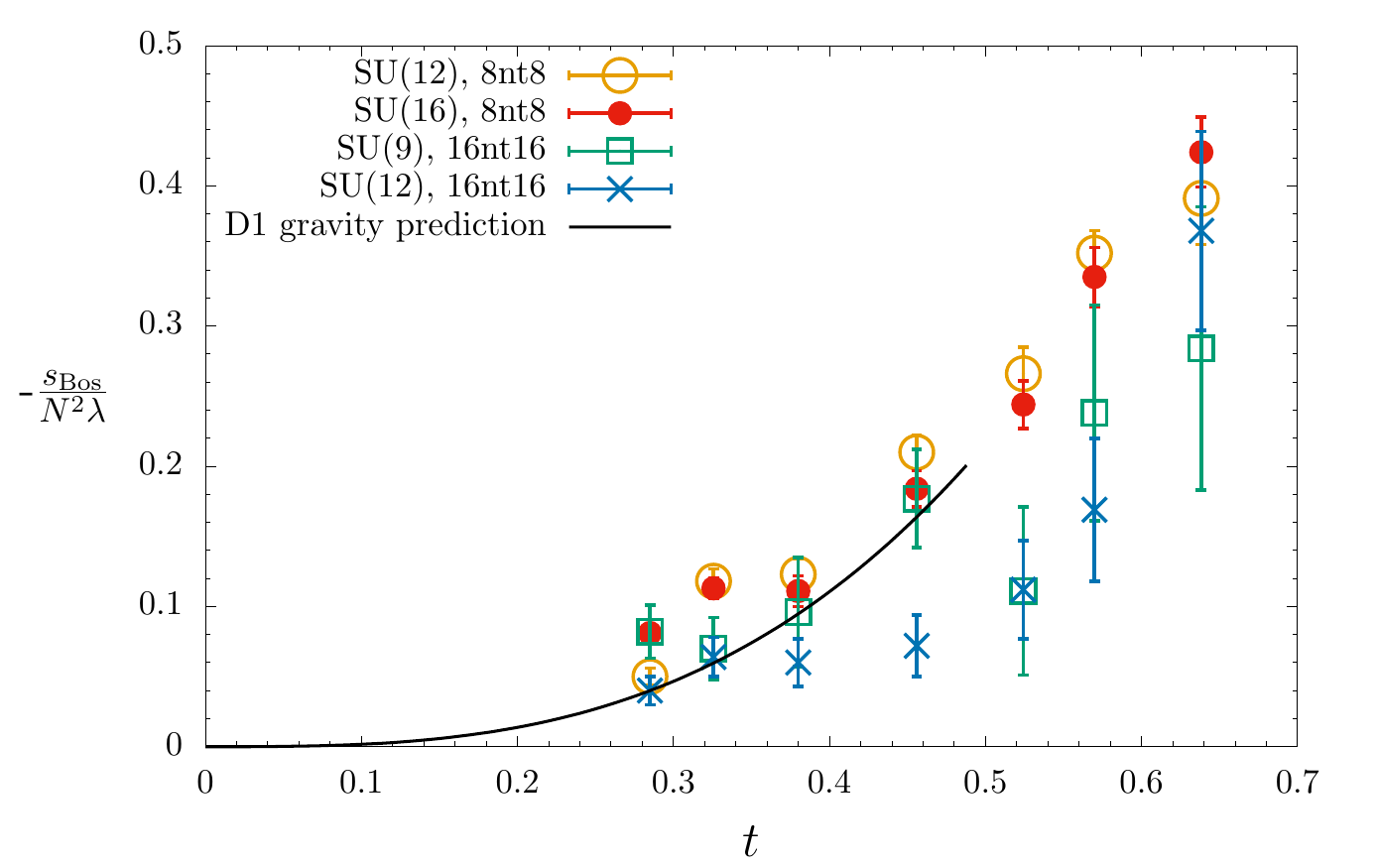}
  \caption{\label{fig:alpha1}Bosonic action density versus dimensionless temperature $t$ for $8\times 8$ and $16\times 16$ lattices (aspect ratio $\al = 1$) with gauge groups SU(9), SU(12) and SU(16).  All points are results of $\mu^2 \to 0$ extrapolations and for low temperatures $t < 0.4$ our results are in reasonable agreement with the D1-phase gravity prediction (solid curve).  The points around $t \simeq 0.5$ show unusual sensitivity to the lattice size and to $N$.  This may be related to the nearby transition around $t_c \simeq 0.47$ (cf.~\protect\fig{fig:crit}), which might lead to underestimated uncertainties.}
\end{figure}

Next, in \fig{fig:alpha1} we show our corresponding bosonic action density results for $\al = 1$ lattice sizes $8\times 8$ and $16\times 16$.
We compare our results to the D1-phase gravity prediction, and see reasonable agreement for $t \lesssim 0.4$.
At slightly larger $t \simeq 0.5$ we observe unusual sensitivity to the lattice size and to $N$.
We suspect that the uncertainties on some of these $\mu^2 \to 0$ extrapolated results may be underestimated.
Figure~\ref{fig:crit} suggests that for $\al = 1$ the deconfinement transition occurs around $t_c \simeq 0.47$, and at the nearby $t \approx 0.46$ we observe unusually long autocorrelations for the SU(12) $16\times 16$ ensembles with larger $\mu^2 \gtrsim 0.003$.
Some of the SU(12) $16\times 16$ $\mu^2 \to 0$ extrapolations around $t \simeq 0.5$ also produce unusually large $2 \lesssim \chidof \lesssim 4$.
Future calculations with larger $N$ and larger lattice sizes, as needed to quantify systematic uncertainties and enable controlled extrapolations to the large-$N$ continuum limit, should clarify this behavior.
Since the deconfinement transition occurs at quite large $t$, we do not expect our results for $t > t_c$ to be well described by the gravity D0-phase behavior, and indeed they are not.
In order to see the gravity D0-phase behavior emerge at low temperature we need to consider smaller $\al < 1$ so that the transition occurs at $t_c \ll 1$.

\subsection{D0-phase thermodynamics}
\begin{figure}[tbp]
  \centering
  \includegraphics[height=\figheight]{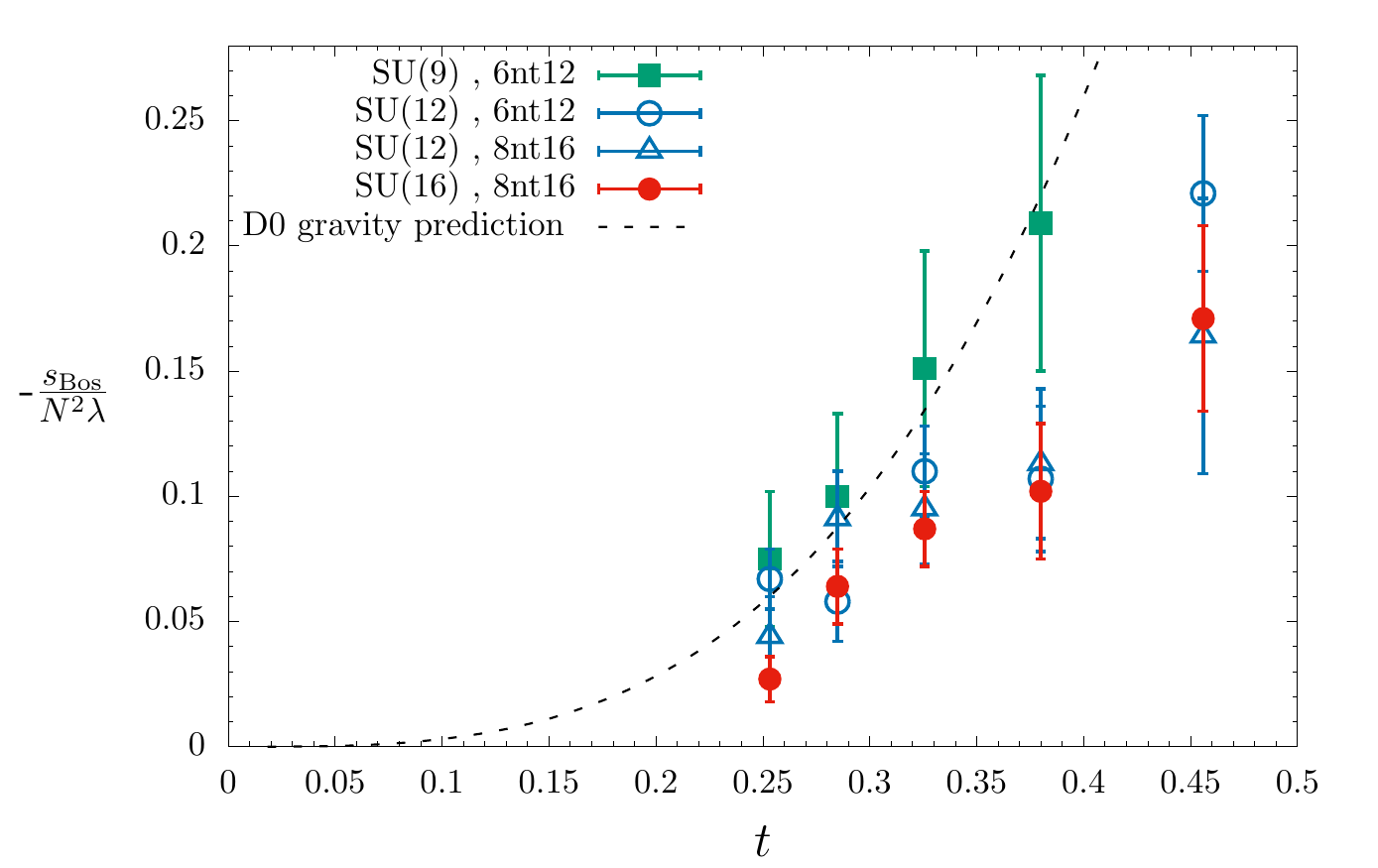}
  \caption{\label{fig:alpha05}Bosonic action density versus dimensionless temperature $t$ for $6\times 12$ and $8\times 16$ lattices (aspect ratio $\al = 1 / 2$) with gauge groups SU(9), SU(12) and SU(16).  All points are results of $\mu^2 \to 0$ extrapolations.  For sufficiently small $t$ our results are consistent with the D0-phase gravity prediction.}
\end{figure}
\begin{figure}[tbp]
  \centering
  \includegraphics[height=\figheight]{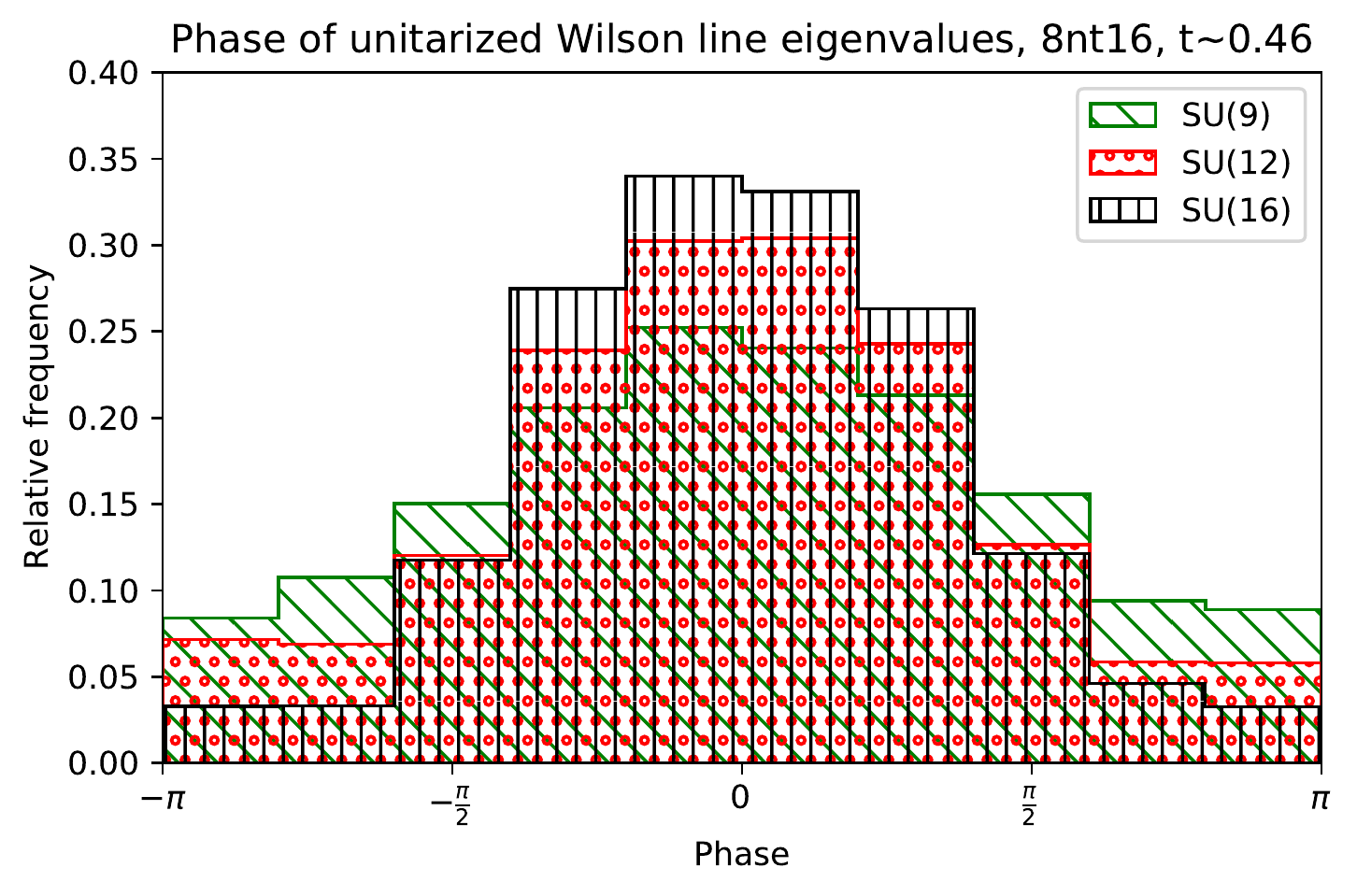}
  \caption{\label{fig:WLeig-al05}Distributions of Wilson line eigenvalue phases, as in \protect{\fig{fig:WLeig-highT}}, for $8\times 16$ lattices at $t \approx 0.46$ with $\mu^2 \approx 0.004$.  The intermediate distributions, which become more compact as $N$ increases, are consistent with expectations from the D0~phase of the gravity dual.}
\end{figure}

The final numerical results we present consider our smallest aspect ratio $\al = 1 / 2$.
Recall from \tab{tab:tori} that this lattice geometry is actually equivalent to a rectangular ($\ga' = 0$) torus with $\al' = 1 / \sqrt{3}$, so that $r_{\beta}' = \sqrt{3} r_{\beta} / 2$.
In \fig{fig:alpha05} we plot the bosonic action density for $N = 9$, 12 and 16 from $6\times 12$ and $8\times 16$ lattices.
Again the points shown are results of $\mu^2 \to 0$ extrapolations.
From \fig{fig:crit} we expect the system to be spatially deconfined for the low temperature range $0.25 \lesssim t \lesssim 0.5$ shown here.
Eventually at very low temperatures, presumably around $t \simeq 0.12$, it should confine, but we are not yet able to probe such a low temperature regime.
The dashed curve is the low-temperature gravity prediction from the (spatially deconfined) D0~phase, Eq.~\eqref{eq:skewD0phase}, which is indeed consistent with the data for $t \lesssim 0.35$.
Figure~\ref{fig:WLeig-al05} shows intermediate distributions for the Wilson line eigenvalue phases on $8\times 16$ lattices at $t \approx 0.46$ (with $\mu^2 \approx 0.004$), which become more compact as $N$ increases.
This behavior supports our conclusion that the system is spatially deconfined in this region of the phase diagram, consistent with the dual gravity approaching the D0~phase in the large-$N$ limit over this temperature range.

\section{Conclusions}
We have studied two-dimensional SYM with maximal supersymmetry compactified on a flat but skewed torus in which an anti-periodic boundary condition is imposed on the fermion fields wrapping one of the cycles.
The theory contains three dimensionless parameters: $r_L$, $r_{\beta} = 1 / t$ and the skewing angle $\cos\theta = \ga$.
From the holographic conjecture, at low `generalized temperature' $t \ll 1$ this theory should give a description of a dual gravitational system containing various types of black holes arising in Type~IIA and IIB supergravity.
The phase diagram of the gravitational system is expected to contain a region where homogeneous D1 (black string) solutions dominate and another in which localized D0 (black hole) solutions dominate.
The critical line separating these two regions in the dual gravitational system is conjectured to be dual to a first-order deconfinement transition with the spatial Wilson loop magnitude $P_L$ serving as an order parameter.

We use lattice gauge theory to explore and test this holographic conjecture using a recently constructed lattice action based on a formalism that maintains an exact supersymmetry at non-zero lattice spacing.
The construction singles out a particular skewing angle $\ga = -1 / 2$, which allows us to test holography both for the usual rectangular tori and also---for the first time---for skewed tori as well.

We have mapped out the phase diagram of the SYM system and indeed find a line of transitions separating a spatially confined phase from a deconfined one.
The parametric form of this phase boundary agrees with the results from the gravity dual.
Furthermore, the action density computed in either phase is consistent at low temperatures with the corresponding black hole thermodynamics.
Our results can be seen as the first step in checking the predictions of gauge/gravity duality in a situation that is distinct from SYM quantum mechanics and has a more subtle phase structure.
However we expect it will be a considerable technical challenge to reach for this two-dimensional theory the degree of precision that is now state of the art in the quantum mechanical case, featuring fully quantified systematic uncertainties and controlled extrapolations to the large-$N$ continuum limit.

\subsection*{Acknowledgements}
We thank Krai Cheamsawat, Joel Giedt, Anosh Joseph and Jamie Hudspith for helpful conversations.
RGJ and TW thank the organizers of the April 2017 ``Quantum Gravity, String Theory and Holography'' workshop at Kyoto University's Yukawa Institute for Theoretical Physics, where this work was first presented and benefited from interesting discussions.
This work was supported by the U.S.~Department of Energy (DOE), Office of Science, Office of High Energy Physics, under Award Numbers DE-SC0008669 (DS) and DE-SC0009998 (SC, RGJ, DS). 
Numerical calculations were carried out on the HEP-TH cluster at the University of Colorado, on the DOE-funded USQCD facilities at Fermilab, and at the San Diego Computing Center through XSEDE supported by National Science Foundation Grant No.~ACI-1053575.

\appendix
\section{\label{app:modular}Modular group and fundamental domain}
Let us recall some facts about the usual modular group $G_{\text{std}}$ and its action on the complex torus parameter $\uptau \in H$, with $H$ the upper half complex plane excluding the real line (so that $\mathrm{Im}(\uptau) > 0$).
The action is given by
\begin{equation}
  \uptau' = \frac{a \uptau + b}{c \uptau + d}
\end{equation}
where $a, b, c, d \in \Zbb$ and $ad - bc = 1$, which corresponds to an element $\left(\begin{array}{cc} a & b \\ c & d \end{array}\right) \in \text{SL}(2, \Zbb)$.
This group is generated by $S$, $T$ where $S(z) = -1 / z$ and $T(z) = z + 1$.
The fundamental domain for this action on $\uptau$ is
\begin{equation}
  D_{\text{std}} = \left\{\uptau \big| 1 \le |\uptau|, \; - \frac{1}{2} \le \mathrm{Re}(\uptau) \le \frac{1}{2} \right\}.
\end{equation}
In fact we may also take the generators of the group to be $R$ and $T$, where $R(z) = z / (z + 1)$, since $S = T^{-1} R T^{-1}$.
These generators $R$, $T$ are associated to the SL$(2, \Zbb)$ matrices $\left(\begin{array}{cc} 1 & 0 \\ 1 & 1 \end{array}\right)$ and $\left(\begin{array}{cc} 1 & 1 \\ 0 & 1 \end{array}\right)$, respectively.

In this work we are interested in the subset $G$ of the modular group $G_{\text{std}}$ that leaves our fermion boundary conditions invariant, namely the above with $a \in 2\Zbb$, $b \in 2\Zbb - 1$ and $c \in \Zbb$.
It is easy to see that $G$ is a subgroup of $G_{\text{std}}$.
The fundamental domain for the action of this new group, $G$, on $H$ is
\begin{equation}
  D = \left\{\uptau\big|1 \le |\uptau \pm 1|, \; -1 \le \mathrm{Re}(\uptau) \le 1\right\}.
\end{equation}
It is generated by the transformations $R(z)$ and $U(z) = T^2(z) = z + 2$, with $R$ as above and $U$ corresponding to the SL$(2, \Zbb)$ matrices $\left(\begin{array}{cc} 1 & 2 \\ 0 & 1 \end{array}\right)$.
We now prove these statements using a simple adaptation of Serre's arguments concerning the domain and generators of the usual modular group~\cite{Serre}.

Following Serre we firstly consider the subgroup $G'$ of $G$ generated by $R$ and $U$, and later show this is in fact the group $G$. \\

\noindent \textit{\underline{Proposition}:} For any $z \in H$ there exists some $g \in G'$ such that $g z \in D$.

\begin{proof}
  Let $z \in H$ and $g$ be an element of the group $G'$ (which is a subgroup of $G$).
  We note that
  \begin{equation}
    \mathrm{Im}(g z) = \frac{\mathrm{Im}(z)}{|c z + d|^2}
  \end{equation}
  and for integers $c, d$ this implies that there exists a $g$ which maximizes $\mathrm{Im}(g z)$.
  Taking such a $g$, then we may choose an integer $n$ so that $z' = U^n g z$ has real part between $\pm 1$.
  In fact $z' \in D$ as we may see by considering $R z'$ and $R^{-1} z'$.
  For any $w \in H$ we have
  \begin{align}
    \mathrm{Im}(R w) & = \frac{\mathrm{Im}(w)}{|w + 1|^2} &
    \mathrm{Im}(R^{-1} w) & = \frac{\mathrm{Im}(w)}{|w - 1|^2}.
  \end{align}
  Thus if $z' \not\in D$, so that either $|z' + 1| < 1$ or $|z' - 1| < 1$, then $g' = R U^n g$ and $g'' = R^{-1} U^n g$ are also elements of $G'$, but either $\mathrm{Im}(g' z) > \mathrm{Im}(g z)$ or $\mathrm{Im}(g'' z) > \mathrm{Im}(g z)$.
  However $\mathrm{Im}(g z)$ was assumed to be maximal, hence we conclude that $z' \in D$.
\end{proof}

\noindent \textit{\underline{Proposition}:} Given two distinct points $z, z' \in D$, then there exists $g \in G$ so that $z' = g z$ only for $z, z' \in \partial D$ (i.e., in the boundary of the fundamental domain).

\begin{proof}
  From the usual arguments about the modular group, we know that for $z \in D_{\text{std}}$ and distinct $z' \in D$ so that $z' = g z$ for $g \in G_{\text{std}}$, then $g = S$, $T$ or $T^{-1}$.
  Hence for distinct $z, z' \in D$ such that $z' = g z$ for $g \in G$ then $g$ is one of $\{S, S^{-1}, T, T^{-1}, T^2, T^{-2}, S T, S T^{-1}, T S, \\ T^{-1} S\}$ (noting that $S^2 = 1$).
  Considering the corresponding SL$(2, \Zbb)$ matrices, we see only the elements $T^2$ and $T^{-2}$ from this list can be elements of $G$, being $U$ and $U^{-1}$, respectively.
  However the only distinct points $z, z' \in D$ related by $g = U$ or $U^{-1}$ are those on the boundaries $\mathrm{Re}(z) = \pm 1$, $\mathrm{Re}(z') = \mp 1$.
\end{proof}

\noindent \textit{\underline{Proposition}:} The subgroup $G'$ is in fact the full group, so $G' = G$.

\begin{proof}
  Consider $z_0 = 2i$ so that $z_0$ is in the interior of $D$.
  Then choose any $g \in G$ and construct $z = g z_0$.
  However from the first proposition above there exists some $g' \in G'$ such that $z' = g' g z_0 \in D$.
  Thus we have $z_0, z' \in D$, with $z_0 \not\in \partial D$ and $z' = h z_0$ for $h = g' g \in G$.
  However from the above proposition that can only be true for $h = 1$, and hence $g = g'^{-1}$.
  Hence $g \in G'$, and thus $G' = G$.
\end{proof}

A useful corollary of the above construction simply follows that will have physical importance for us in the main text. \\

\noindent \textit{\underline{Corollary}:} Given $\uptau \in D$, then for any $g \in G$ we have $\mathrm{Im}(g \uptau) \le \mathrm{Im}(\uptau)$.

\section{\label{app:scalar}Scalar example: an $A_4^*$ lattice and its dimensional reduction}
Here we consider a scalar field theory in four dimensions discretized on an $A_4^*$ lattice.
We use this to illustrate explicitly the reduction of such a lattice theory to a two-dimensional theory on an $A_2^*$ lattice.
We take a discretization analogous to that we use for $\cN = 4$ SYM, having the lattice action
\begin{equation}
  \label{eq:latscalar}
  S_{\text{lat}} = \frac{1}{\lalat} \sum_{\vn \in \Zbb^4} \left[\sum_{a = 1}^5 \left[\cD_a \phi(\vn)\right]^2 + \phi(\vn)^4 \right],
\end{equation}
with the lattice derivative taken as
\begin{equation}
  \cD_a \phi(\vn) = \phi\left(\vn + \hatbmu_{(a)}\right) - \phi(\vn).
\end{equation}
The lattice variable $\phi$ lives at lattice sites $\vn \in \Zbb^4$ and we have included an interaction term and coupling, normalized in analogy with a gauge coupling.
Here the vectors $\hatbmu_{(a)}$ have components $\hat{\mu}_{(\nu)}^{\al} = \delta_{\nu}^{\al}$ for $\al, \nu = 1, \ldots, 4$ and $\hat{\mu}_{(5)}^{\al} = (-1, -1, -1, -1)$.
The kinetic term is differenced symmetrically in these five directions.

Consider continuum coordinates $y^{\mu} = \Delta\, n^{\mu}$ with $\Delta$ the scale setting the lattice size (proportional to the lattice spacing).
Taking the continuum limit $\Delta \to 0$ and assuming a suitably smooth scalar field $\phi$ we may expand
\begin{equation}
  \phi\left(\vn + \hatbmu_{(a)}\right) - \phi(\vn) = \Delta\, \hat{\mu}_{(a)}^{\nu} \left. \pderiv{\phi}{y^{\nu}}\right|_{\vy = \Delta\, \vn}.
\end{equation}
Then, using $\sum_{\vn \in \Zbb^4} f(\vn) \simeq \frac{1}{\Delta^4} \int d^4 y\, f(y)$ for a (suitably smooth) function $f$ in the $\Delta \to 0$ limit, and defining a continuum field $\Phi = \frac{1}{\Delta} \phi$, the lattice action has the continuum limit $S_{\text{lat}} \to S_{\text{4-cont}}$, where
\begin{equation}
  \label{eq:4dact1}
  S_{\text{4-cont}} = \frac{1}{\lalat} \int d^4 y \left[\sum_{\mu = 1}^4 \left(\pderiv{\Phi}{y^{\mu}}\right)^2 + \left(\sum_{\mu = 1}^4 \pderiv{\Phi}{y^{\mu}}\right)^2 + \Phi^4 \right] = \frac{1}{\lam_4} \int d^4 y \sqrt{|g|} \Big[g^{\mu\nu} \partial_{\mu} \Phi \partial_{\nu} \Phi + \Phi^4 \Big].
\end{equation}
Here the components of the metric are $g_{\mu\nu} = \delta_{\mu\nu} - \frac{1}{5}$, and $|g| = \det{g_{\mu\nu}} = \frac{1}{5}$.
The continuum coupling $\lam_4$ is related to the lattice coupling by $\lam_4 = \lalat / \sqrt{5}$.
We may change to canonical flat-space coordinates $x^{\al} = y^{\mu} e_{(\mu)}^{\al}$, where
\begin{equation}
  e_{(\mu)}^{\al} = \left(\begin{array}{cccc}
     \frac{1}{\sqrt{2}} &  \frac{1}{\sqrt{6}} &  \frac{1}{\sqrt{12}} & \frac{1}{\sqrt{20}} \\
    -\frac{1}{\sqrt{2}} &  \frac{1}{\sqrt{6}} &  \frac{1}{\sqrt{12}} & \frac{1}{\sqrt{20}} \\
     0                  & -\frac{2}{\sqrt{6}} &  \frac{1}{\sqrt{12}} & \frac{1}{\sqrt{20}} \\
     0                  &  0                  & -\frac{3}{\sqrt{12}} & \frac{1}{\sqrt{20}}
  \end{array}\right).
\end{equation}
Then
\begin{equation}
  \label{eq:4dact2}
  S_{\text{4-cont}} = \frac{1}{\lam_4} \int d^4 x \left[\left(\partial_{\al} \Phi\right)^2 + \Phi^4 \right],
\end{equation}
and the lattice sites are located at $x^{\al} = \Delta\, n^{\mu} e_{(\mu)}^{\al}$ for $\vn \in \Zbb^4$.
Recognizing these $\ve_{(\mu)}$ as basis vectors for the $A_4^*$ lattice, and noting that Eq.~\eqref{eq:latscalar} includes a difference also in the direction $\ve_{(5)} = -\sum_{\mu = 1}^4 \ve_{(\mu)}$, we see our original lattice theory is indeed defined on an $A_4^*$ lattice.
While from the explicit coordinate presentation above it isn't obvious, this lattice is maximally symmetric as we should expect from Eq.~\eqref{eq:latscalar},
\begin{align}
  \ve_{(a)} \cdot \ve_{(b)} & = \delta_{ab} - \frac{1}{5} &
  a, b & \in 1, \ldots, 5,
\end{align}
and the lattice spacing $a_4$ along all five directions $\ve_{(a)}$ is $a_4 = \Delta \sqrt{4 / 5}$.

Now suppose we are interested in reducing this four-dimensional lattice theory to two dimensions.
We take the lattice variables to be independent of the $\ve_{(3, 4)}$ lattice directions, and restrict the lattice sum $\sum_{\vn \in \Zbb^4}$ only to a two-dimensional slice of the original four-dimensional lattice, $\vn = (n^1, n^2, 0, 0)$ with $(n^1, n^2) \in \Zbb^2$.
The lattice action becomes
\begin{equation}
  \label{eq:2dLatticeAct}
  S_{\text{lat}}^{\text{(red)}} = \frac{1}{\lalat} \sum_{(n^1, n^2) \in \Zbb^2} \left[\sum_{a = 1}^2 \left(\phi\left(\vn + \hatbmu_{(a)}\right) - \phi(\vn)\right)^2 + \left(\phi\left(\vn - \hatbmu_{(1)} - \hatbmu_{(2)}\right) - \phi(\vn)\right)^2 + \phi(\vn)^4\right]
\end{equation}
As above, taking continuum coordinates $y^i = \Delta\, n^i$ with $i = 1, 2$, and the continuum limit $\Delta \to 0$, for a suitably smooth function $f$ we have $\sum_{(n^1, n^2) \in \Zbb^2} f(\vn) \simeq \frac{1}{\Delta^2} \int d^2 y\, f(y)$.
The lattice action has continuum limit
\begin{equation}
  \label{eq:2dact1}
  S_{\text{2-cont}} = \frac{\Delta^2}{\lalat} \int d^2 y \left[\sum_{i = 1}^2 \left(\pderiv{\Phi}{y^i}\right)^2 + \left(\sum_{i = 1}^2 \pderiv{\Phi}{y^i}\right)^2 + \Phi^4 \right] = \frac{1}{\lam_2} \int d^2 y \sqrt{|h|} \left[h^{ij} \partial_i \Phi \partial_j \Phi + \Phi^4 \right],
\end{equation}
with the metric components $h_{ij} = \delta_{ij} - \frac{1}{3}$ so that $|h| = \frac{1}{3}$ and the two-dimensional continuum coupling $\lam_2 = \lalat / (\Delta^2 \sqrt{3})$.
We may again move to canonical flat-space coordinates $\tilde{x}^m = y^i \tilde{e}_{(i)}^m$ with $m = 1, 2$, where
\begin{equation}
  \tilde{e}_{(i)}^m = \left(\begin{array}{cc}
     \frac{1}{\sqrt{2}} & \frac{1}{\sqrt{6}} \\
    -\frac{1}{\sqrt{2}} & \frac{1}{\sqrt{6}}
  \end{array}\right).
\end{equation}
Then
\begin{equation}
  \label{eq:2dact2}
  S_{\text{2-cont}} = \frac{1}{\lam_2} \int d^2 \tilde{x} \left[\left(\partial_m \Phi\right)^2 + \Phi^4 \right],
\end{equation}
and the two-dimensional lattice sites are now located at $\tilde{x}^m = \Delta\, n^i \tilde{e}_{(i)}^m$ for $\vn \in \Zbb^2$.\footnote{These coordinate positions are those of the original $A_4^*$ lattice restricted to the 1 and 2~directions, so $\tilde{x}^m = x^m = \Delta\, (n_1, n_2, 0, 0)^{\mu} e_{(\mu)}^m$ for $(n_1, n_2) \in \Zbb^2$.}
Now defining $\tilde{\ve}_{(3)} = -\tilde{\ve}_{(1)} - \tilde{\ve}_{(2)}$, we see the reduced lattice action, Eq.~\eqref{eq:2dact1}, is defined using differences generated by $\left\{\tilde{\ve}_{(1)}, \tilde{\ve}_{(2)}, \tilde{\ve}_{(3)}\right\}$.
We recognize this as an $A_2^*$ lattice action, noting that
\begin{align}
  \tilde{\ve}_{(a)} \cdot \tilde{\ve}_{(b)} & = \delta_{ab} - \frac{1}{3} &
  a, b & \in 1, \ldots, 3,
\end{align}
and the lattice spacing $a_2$ along these three directions $\tilde{\ve}_{(a)}$ is $a_2 = \Delta \sqrt{2 / 3}$.

Finally we note that the two-dimensional continuum action~\eqref{eq:2dact2} is simply the dimensional reduction of the four-dimensional action~\eqref{eq:4dact2}.
More precisely, if we consider the four-dimensional continuum action~\eqref{eq:4dact1} and take the $y^3$ and $y^4$ coordinates to be periodic with period $\Delta$, so that $y^{3, 4} \sim y^{3, 4} + \Delta$, then for $\Delta \to 0$ we may Kaluza--Klein reduce on these directions. 
The zero modes determine the reduced two-dimensional action, which after integrating over the $y^{3, 4}$ directions gives precisely the two-dimensional action~\eqref{eq:2dact1}, with the two- and four-dimensional couplings related by
\begin{equation}
  \label{eq:KKcoupling}
  \lam_2 = \frac{\lam_4}{\Delta^2} \sqrt{\frac{5}{3}}. 
\end{equation}
The factor $\sqrt{5 / 3}$ arises as the Kaluza--Klein reduction is over small periodic directions that are not orthogonal to each other or to the extended $y^i$ directions.

\section{\label{app:num}Numerical details}
We use the standard rational hybrid Monte Carlo (RHMC) algorithm~\cite{Clark:2006fx} implemented in the publicly available parallel software described by \refcite{Schaich:2014pda}.\footnote{{\tt\href{https://github.com/daschaich/susy}{github.com/daschaich/susy}}}
In the course of this work we have improved this software to enable the SU($N$) truncation of the gauge links discussed near the end of \secref{sec:lattice_4d}, as well as to add the scalar potential terms in Eqs.~\eqref{eq:single_trace} and \eqref{eq:center}.
These additions, along with related improvements to the large-$N$ performance of the code and other advances, will soon be presented in another publication~\cite{parallel_imp}.

\begin{figure}[tbp]
  \centering
  \includegraphics[height=\figheight]{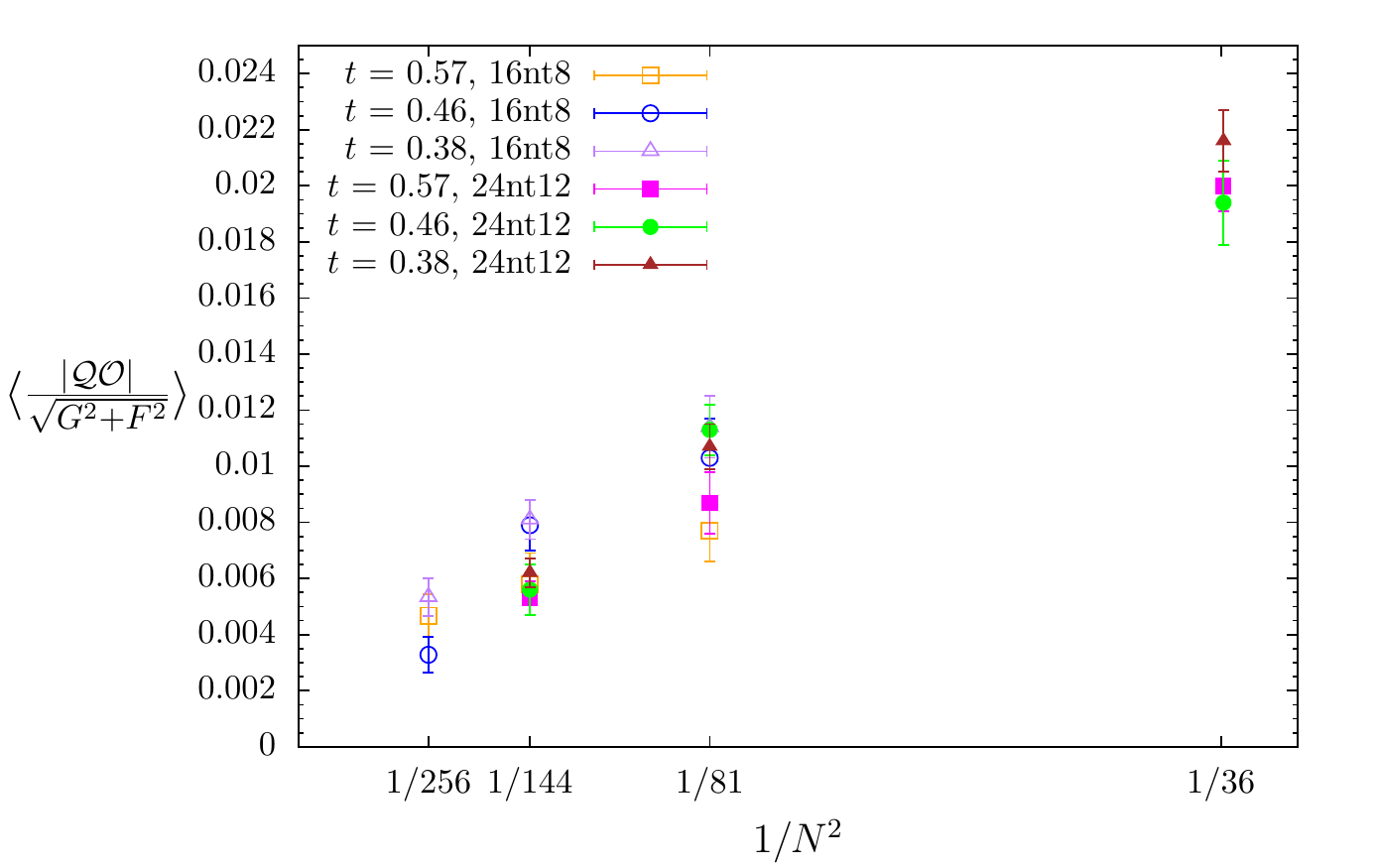}
  \caption{\label{fig:ward}Violations of a \cQ Ward identity vs.\ $1 / N^2$ for $16\times 8$ and $24\times 12$ lattices (aspect ratio $\al = 2$) with fixed $\mu^2 / \lalat = c_W^2 / \lalat = 0.01$.  The violations are suppressed $\sim 1 / N^2$ and show little dependence on the lattice size or the temperature in the range $0.38 \leq t \leq 0.57$.}
\end{figure}
\begin{figure}[tbp]
  \centering
  \includegraphics[height=\figheight]{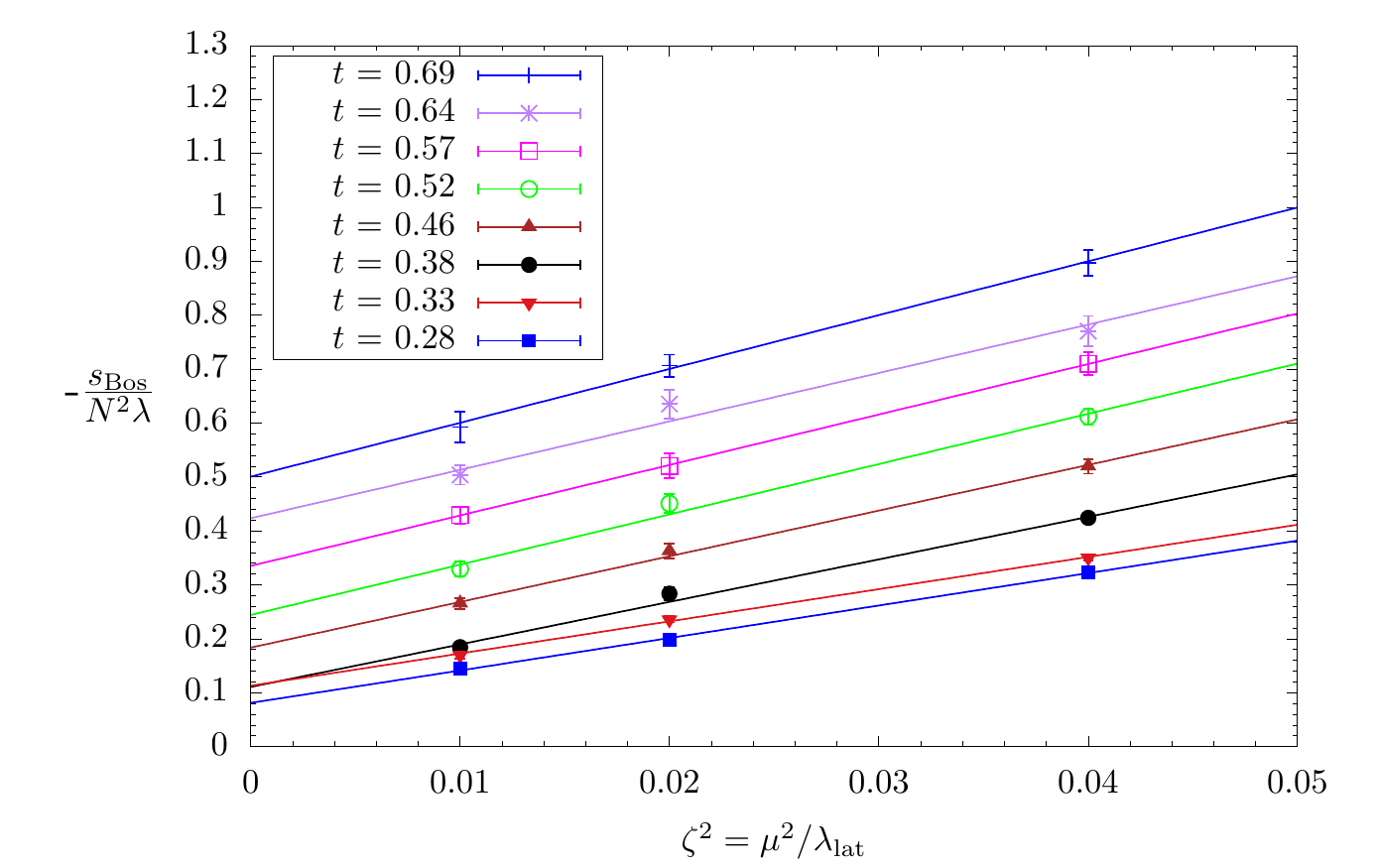}
  \caption{\label{fig:extrap}A representative sample of linear $\mu^2 \to 0$ extrapolations of bosonic action data for SU(16) $8\times 8$ lattices (aspect ratio $\al = 1$).  The intercepts correspond to the red points in \protect\fig{fig:alpha1}.}
\end{figure}

The results presented in the body of this paper involve eight aspect ratios $\al = r_L / r_{\beta} = 1 / 2$, 1, $3 / 2$, 2, $8 / 3$, 4, 6 and 8, investigated for up to five SU($N$) gauge groups with $N = 3$, 6, 9, 12 and 16.
In order to ensure that the soft-$\cQ$-breaking scalar potential terms~\eqref{eq:single_trace} and \eqref{eq:center} introduce only small effects, we require that $\mu^2, c_W^2 \ll \lalat$.
Specifically, as \lalat varies we fix the ratio $\mu^2 / \lalat = 0.01$, 0.02, 0.03 or 0.04, with either $c_W^2 = 0$ or $c_W^2 = \mu^2$.
In \fig{fig:ward} we show violations of a \cQ Ward identity (discussed in detail by Refs.~\cite{Catterall:2014vka, Schaich:2015ppr}), which are small (percent-level at most) and decrease roughly $\propto 1 / N^2$ with fixed $\mu^2 / \lalat = c_W^2 / \lalat = 0.01$.
This establishes that both the scalar potential and the SU($N$) truncation of the gauge links have insignificant numerical effects for sufficiently large $N$.
Finally, in \fig{fig:extrap} we show a representative sample of the linear $\mu^2 \to 0$ extrapolations that produce the results for the bosonic action plotted in Figs.~\ref{fig:alpha2}, \ref{fig:alpha1} and \ref{fig:alpha05}.
The extrapolations shown here, for SU(16) $8\times 8$ lattices, have acceptable $0.01 \leq \chidof \leq 1.95$ and confidence levels $0.94 \geq \text{CL} \geq 0.16$, where
\begin{equation}
  \mbox{CL} = 1 - P(a, x) = \frac{1}{\Gamma(a)} \int_x^{\infty} dt\ e^{-t}\ t^{a - 1},
\end{equation}
$a = \mbox{d.o.f.} / 2$ and $x = \chi^2 / 2$.
Their intercepts correspond to the red points in \fig{fig:alpha1}.

The RHMC algorithm treats the factor of $e^{-S}$ in the partition function~\eqref{eq:partfunc} as a Boltzmann weight, requiring that the Euclidean action $S$ be real and non-negative.
However, gaussian integration over the fermion fields of $\cN = (8, 8)$ SYM produces a pfaffian that is potentially complex,
\begin{equation}
  \int \left[d\Psi\right] e^{-\Psi^T \cD \Psi} \propto \pf \cD = |\pf \cD| e^{i\phi},
\end{equation}
where \cD is the fermion operator.
Therefore all our numerical calculations `quench' the phase $e^{i\phi} \to 1$~\cite{Schaich:2014pda}.
In principle, the true expectation values $\vev{\cO}$ can be recovered from phase-quenched (`$_{pq}$') calculations via reweighting,
\begin{equation}
  \vev{\cO} = \frac{\vev{\cO e^{i\phi}}_{pq}}{\vev{e^{i\phi}}_{pq}}
\end{equation}
\begin{align}
  \vev{\cO}_{pq} & = \frac{\int[d\cU] \ \cO e^{-S_B}\ |\pf D|}{\int[d\cU] \ e^{-S_B} \ |\pf D|} &
  \vev{\cO} = \frac{\int[d\cU] \ \cO e^{-S_B}\ \pf D}{\int[d\cU] \ e^{-S_B} \ \pf D}.
\end{align}
Reweighting requires measuring the pfaffian phase $\vev{e^{i\phi}}_{pq}$, and fails if this expectation value is consistent with zero.

\begin{table}[htbp]
  \centering
  \renewcommand\arraystretch{1.2}   
  \addtolength{\tabcolsep}{3 pt}    
  \begin{tabular}{c|c|c|c}
    \hline
    $\al = N_x / N_t$ & $N_x \times N_t$ & $t$  & $1 - \vev{\cos\phi}$     \\
    \hline
    4                 & $16\times 4$     & 4.56 & $37(17)\times 10^{-12}$  \\ 
    \hline
    $3 / 2$           & $12\times 8$     & 0.76 & $55(18)\times 10^{-7}$   \\ 
    \hline
    1                 & $8\times 8$      & 0.38 & $49(17)\times 10^{-7}$   \\ 
                      &                  & 0.76 & $2.93(77)\times 10^{-8}$ \\ 
    \hline
  \end{tabular}
  \caption{\label{tab:pfaffian}Tests of pfaffian phase fluctuations for SU(3) ensembles, considering three aspect ratios \al and a range of temperatures $0.38 \leq t \leq 4.56$.  In all cases $1 - \vev{\cos\phi} \ll 1$ corresponds to very small fluctuations in the phase itself, $\vev{e^{i\phi}}_{pq} \approx 1$ so that phase reweighting has no practical effect.}
\end{table}

Previous lattice studies of $\cN = (2, 2)$ and $\cN = (8, 8)$ SYM theories in two dimensions found $\vev{e^{i\phi}}_{pq} \approx 1$ even at non-zero lattice spacing, with deviations from unity vanishing rapidly upon approach to the continuum limit~\cite{Hanada:2010qg, Catterall:2011aa, Mehta:2011ud, Galvez:2012sv}.
Since we use a different lattice action than those considered previously, for a few SU(3) ensembles we have checked that this remains true in our current work.
Table~\ref{tab:pfaffian} collects the results of these tests, considering $1 - \vev{\cos\phi}$ to ensure that positive and negative fluctuations in $\phi$ cannot cancel out.
In all cases we find $1 - \vev{\cos\phi} \ll 1$, corresponding to $\vev{e^{i\phi}}_{pq}$ close enough to unity that reweighting has no practical effect.

\bibliographystyle{utphys}
\bibliography{2dSYM}
\end{document}